\documentclass[fleqn,usenatbib,useAMS]{mnras}

\usepackage{graphicx}	
\usepackage{amsmath}	
\usepackage{multicol}        
\usepackage{bm}		
\usepackage{pdflscape}	

\usepackage[T1]{fontenc}
\usepackage{ae,aecompl}

\usepackage{newtxtext,newtxmath}

\title[Mismatched mirror images]{Mirror images of lensed star clusters with mismatched spectral energy distributions: A possible signature of top-heavy stellar initial mass functions and extreme stars in high-redshift star clusters}

\author[E. Zackrisson]{Erik Zackrisson$^{1}$
\thanks{Contact e-mail: \href{mailto:erik.zackrisson@physics.uu.se}{erik.zackrisson@physics.uu.se}}, Jose M. Diego,$^{2}$, Jose M. Palencia$^{2}$, Francesco Gabrielli$^{1}$, Armin Nabizadeh$^{1}$,\newauthor Angela Adamo$^{3}$ and Guglielmo Costa$^{4,5}$
\\
$^{1}$Observational Astrophysics, Department of Physics and Astronomy, Uppsala University, Box 524, SE-751 20 Uppsala, Sweden\\
$^{2}$ Instituto de Física de Cantabria (CSIC-UC), Avda. Los Castros s/n, 39005 Santander, Spain\\
$^{3}$ The Oskar Klein Centre, Department of Astronomy, Stockholm University, AlbaNova, SE-10691 Stockholm, Sweden\\
$^{4}$ Dipartimento di Fisica e Astronomia, Università degli studi di Padova, Vicolo dell’Osservatorio 3, Padova, Italy\\
$^{5}$ INAF - Osservatorio Astronomico di Padova, Vicolo dell’Osservatorio 5, Padova, Italy\\
}

\date{Last updated 2020 June 10; in original form 2013 September 5}

\pubyear{2026}

\begin{document}
\label{firstpage}
\pagerange{\pageref{firstpage}--\pageref{lastpage}}
\maketitle

\begin{abstract}
Strongly lensed star clusters have recently been detected up to redshift $z\approx 10$ in galaxy cluster fields using the James Webb Space Telescope (JWST). When pairs of mirror images of such star clusters appear across the lensing critical curve, it is usually assumed that both images will display identical spectral energy distributions (SEDs). However, this assumption may be invalidated in the presence of gravitational microlensing from stars or other compact objects in the lens, since microlensing will affect the SED contribution from bright stars within the star cluster independently in the two mirror images. Here, we explore under what circumstances mismatched mirror-image SEDs are likely to be observable, and argue that SED differences detectable in JWST observations of lensing-cluster fields will be limited to star clusters of mass $< 10^5\ M_\odot$ and ages $\lesssim 5$ Myr. The probability of severely mismatched mirror-image SEDs increases if the stellar initial mass function is very top-heavy and extends to stellar masses $\gg 100\ M_\odot$, as has been suggested to be the case for Population III stars. The prevalence of lensed star clusters with highly discrepant mirror-image SEDs could therefore serve as a probe of very massive stars and extreme stellar populations in the early Universe.
\end{abstract}





\section{Introduction}
Strong gravitational lensing by galaxy clusters at low-to-intermediate redshifts can boost the apparent fluxes of high-redshift background galaxies and stretch such objects into gravitational arcs, thereby allowing their constituents to be partially resolved. In recent years, both lensed star clusters at redshifts as high as $z\approx 6$--10 \citep[e.g.][]{Vanzella23,Adamo24,Mowla24,Messa25} and candidates for individual lensed stars at redshifts up to $z\approx 5$--6 \citep{Welch22,Meena23,Furtak24} have been identified in observations of such arcs\footnote{In some cases, the distinction between a lensed star and a lensed star cluster may be ambiguous, especially when the lensed-star candidate does not show any signs of variability. We note that it has been argued that some of the higher-redshift lensed-star candidates, most notably Earendel at $z\approx 6$, may actually be lensed star clusters \citep{Ji25,Scofield25,Pascale25} rather than pairs of stars as in the \citet{Welch22} Earendel solution.}. 

While the {\it macrolensing} caused by the large-scale gravitational potential of the foreground galaxy cluster is responsible for the formation of these arcs, gravitational {\it microlensing} by stars or other compact objects in the galaxy cluster \citep[e.g.][]{Diego18,Palencia24} may also affect intrinsically small light sources present within these arcs. Microlensing can cause temporal variability in the apparent fluxes of lensed stars \citep[e.g.][]{Miralda-Escude91,Diego18} and -- to a smaller degree - in the integrated light from whole star clusters \citep{Dai20,Dai21}. Here, we explore a different potential effect of microlensing on lensed star clusters: the appearance of multiple images with discrepant spectral energy distributions (SEDs) in single-epoch observations of strong-lensing galaxy cluster fields.

\begin{figure}
\includegraphics[width=\columnwidth]{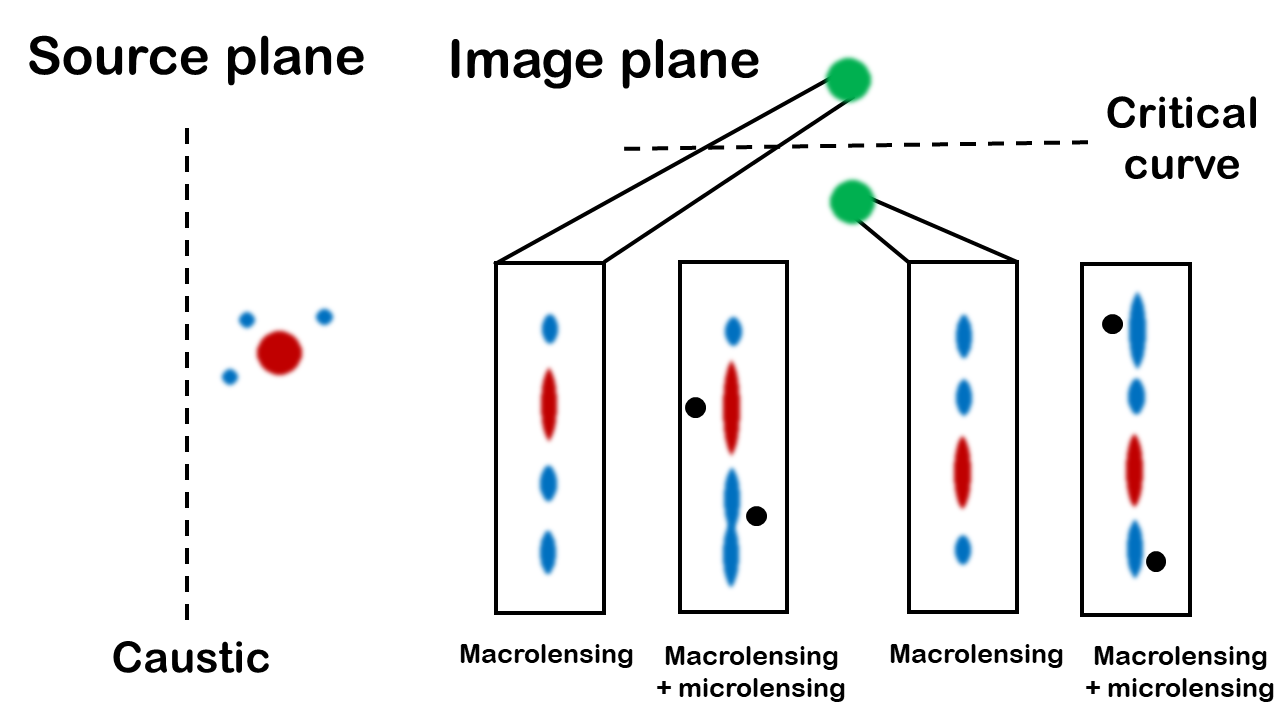}
\label{fig:schematic}
\caption{Schematic illustration of the mechanism by which mirror images of star cluster may develop mismatched SEDs. {\bf Left:} A star cluster featuring a small number of red, evolved stars (here just one; red circle) and a larger number of blue stars (for illustrative purposes just three; blue circles) is located close to the macrolensing caustic in the source plane. {\bf Right:} In the image plane, the light from the blue and red stars blends into two unresolved star cluster macroimages (green circles) located symmetrically on either side of the critical curve. In the case of macrolensing only, the two star cluster images contain the highly stretched images of each star, appearing as mirror-flipped versions of each other. Microlensing by stars or other compact objects (black circles) located in the galaxy cluster that acts as the macrolens distort the magnification of individual stars within the two star cluster images. Due to the separation of the star cluster macroimages in the image plane (typically $\approx 0.1$--1 arcsec), different compact objects microlens the two sets of stellar images independently, which can alter the balance of red to blue light between the two unresolved star cluster images.}
\end{figure}

In the situation where the macrolensing critical curve crosses a gravitational arc, star clusters and star-forming clumps in the lensed galaxy may appear as symmetrically placed ``mirror images'' (counter images) with opposite parity across the critical curve, as seen in recent James Webb Space Telescope (hereafter JWST) observations \citep[e.g.][]{Vanzella23,Adamo24,Abdurrouf25}. While small-scale lensing perturbations from nearby galaxies and dark matter subhalos may cause these mirror images to exhibit different overall magnification and hence different apparent fluxes, the shapes of their SEDs are usually assumed to be identical, since they represent images of the same object. However, the SED of a star cluster may at certain ages and in specific wavelength intervals become dominated by only a small number of stars. Since the microlensing magnification/demagnification of these stars will happen independently in the two mirror images, the integrated SEDs of the mirror images could potentially differ at specific points in time. This lensing situation is schematically illustrated in Figure~\ref{fig:schematic}. Such mirror-image SED discrepancies are temporary and are expected to change as stars in these star clusters move across the microlensing caustic network on timescales of hours to weeks \citep{Dai21}. For very massive star clusters ($\sim 10^7\ M_\odot$), temporal flux variations of single images are at the percentage level \citep{Dai21} and would require observationally expensive monitoring programs with high photometric precision to be detected. For lower-mass star clusters (containing much fewer stars), both the temporal fluctuation amplitudes \citep{Dai20} within individual images and the probability of catching mirror images with discrepant SEDs in single-epoch observations are expected to be higher. Here, we explore under what conditions the latter phenomenon is likely to be detectable. If cases can be found where mirror images exhibit substantially mismatched SEDs, this would indicate not only that microlensing is substantially affecting the images, but also that the stellar content of the star clusters is conducive for this effect to arise. As we will demonstrate, young star clusters with top-heavy stellar initial mass functions (IMFs) and extremely massive stars may be exceptionally prone to exhibit this phenomenon. As it has been argued that the IMF may become more top-heavy at high redshift and/or very low-metallicity \citep[e.g.][]{Chon21,Steinhardt23,Liu24}, a search for mirror images with substantially mismatched SEDs could potentially be used to identify candidates for star clusters with extreme IMFs in the early Universe, such as clusters of chemically pristine Population III (hereafter Pop III) stars.

In Section~\ref{sec:computational machinery}, we describe the computational machinery used to explore the phenomenology of lensed star clusters close to lensing caustics in the presence of microlensing. Section~\ref{sec:normal star cluster} explains the mechanism that gives rise to star cluster mirror images with discrepant SEDs, and investigates the magnitude and prevalence of this effect in the case of a standard IMF for the star cluster. In Section~\ref{sec:extreme IMF} we explore the corresponding effect in the case of extremely top-heavy IMFs and extremely massive stars ($\gg 100\ M_\odot$). Section~\ref{sec:discussion} discusses various other physical properties of high-redshift star clusters, and the overall lensing situation, which could influence the probability of detecting discrepant mirror-image SEDs. Section~\ref{sec:summary} summarizes our findings. 

Throughout this paper, proper distances (not comoving distances) are used as the default to indicate length scales in the high-redshift Universe. A $\Lambda$CDM cosmology with $\Omega_\mathrm{M}=0.3$ and $\Omega_\Lambda=0.7$ and $H_0=70$ km s$^{-1}$ Mpc$^{-1}$ is assumed.

\section{Computational machinery}
\label{sec:computational machinery}
Here, we present the numerical model used to predict the magnification of individual stars within high-redshift star clusters subject to strong lensing by a galaxy cluster lens, and the associated impact on the overall SEDs of star cluster macroimages.

Our model works on a star-by-star basis by treating the magnification of each star in the source plane separately (Section~\ref{subsec:lensing}). We first specify a total stellar mass $M_\mathrm{tot}$ for the star cluster, a stellar initial mass function (IMF) of its constituent stars (Section~\ref{subsec:IMF}), and a spatial distribution of these stars within the star cluster (Section~\ref{subsec:spatial distribution}). After randomly sampling the IMF until the total mass reaches $M_\mathrm{tot}$, the stars are distributed onto the source plane according to the projected spatial distribution. This results in Cartesian coordinates $x$ and $y$ around the centre of the stellar distribution for each star. By adopting an age and metallicity for the star cluster, intrinsic stellar SEDs (Section~\ref{subsec:SED models}) are then assigned to each star. After specifying a distance of the star cluster centre to the macrocaustic in the source plane, the combined macro- and microlensing lensing effects are computed in the image plane and the SEDs of lensed stars within each macroimage are summed up.

\subsection{Macrolensing and microlensing}
\label{subsec:lensing}
We assume the lensing caustic in the source plane at redshift $z_\mathrm{s}$ to be a straight line extended in the $y$ direction, at a distance $d_\mathrm{caustic}$ in units of pc along the $x$ direction from the centre of the star cluster. This implies a distance from the caustic of each star $i$ in the ensemble given by
\begin{equation}
d_{i,\mathrm{caustic}}=d_\mathrm{caustic}+x_\mathrm{i}
\end{equation}

The linear $d_{i,\mathrm{caustic}}$ coordinate is converted into an angular source plane coordinate given by:
\begin{equation}
\theta_i=\frac{1.296\times 10^6\ d_{i,\mathrm{caustic}}} {2\pi D_\mathrm{A}(z_\mathrm{s})}
\label{eq: Theta}
\end{equation}
where $\theta_i$ is given in arcseconds and $D_\mathrm{A}(z_\mathrm{s})$ is the angular-size distance from the observer to the source plane in pc.

The total macrolensing magnification of each star is then derived from the relation for fold caustics:
\begin{equation}
\mu_{\mathrm{macro,tot},i}=\frac{B_0}{\sqrt{\theta_i}},
\label{eq: mu macro tot}
\end{equation}
where $B_0$ is a parameter that typically lies in the range $\approx 10-20$ arcsec$^{1/2}$ for some of the most well-studied strong-lensing clusters \citep{Windhorst18}. If $\theta_i<0$, i.e. if the star has crossed the caustic, then it is not considered further, since the magnification would drop considerably.

In the image plane, each star at $\theta_i>0$ is assumed to display two bright macroimages symmetrically positioned along the lensed arc, with identical radial and tangential macromagnification components $\mu_\mathrm{r}$ and $\mu_\mathrm{t}$. If we furthermore assume that $\mu_\mathrm{r}$ is constant for all stars along the arc (and $\mu_\mathrm{r}\ll \mu_\mathrm{t}$, with $\mu_\mathrm{r}$ typically $\approx 1$--3), $\mu_\mathrm{t}$ of these two images of each star can be expressed as:
\begin{equation}
\mu_{\mathrm{t},i} = \frac{\mu_{\mathrm{macro,tot},i} }{2\mu_\mathrm{r}}
\label{eq:mu tangential}
\end{equation}

The image falling inside the critical curve in the image plane corresponds to negative parity, whereas the one falling outside corresponds to positive parity. 

To determine the brightness of each of the two macroimages in the light of microlensing by intracluster stars in the galaxy-cluster lens at redshift $z_\mathrm{l}$, we make use of the M{\_}SMiLe package \citep{Palencia24}. Based on the adopted $\mu_r$, $\mu_t$ (determined from the distance of each star from the caustic via eqs.~\ref{eq: Theta} ~\ref{eq: mu macro tot} and ~\ref{eq:mu tangential}), and an assumed surface mass density in microlenses $\Sigma_\star$ (in units of $M_\odot$ pc$^{-2}$) in the lens plane, M{\_}SMiLe predicts the magnification probability distribution for the positive and negative images separately. By randomly sampling these distributions, we draw two separate magnifications $\mu_{+}$ and $\mu_{-}$ for each star and determine the corresponding brightness $m_{\mathrm{AB},j,\pm}$ of the $\pm$ parity images in each JWST/NIRCam filter $j$ according to:
\begin{equation}
m_{\mathrm{AB},j,\pm} = m_{\mathrm{AB,intrinsic},j} -2.5\log_{10} \mu_\pm,  
\end{equation}
where $m_{\mathrm{AB,intrinsic},j}$ is the unlensed apparent AB magnitude of the star at redshift $z_\mathrm{s}$ predicted by the stellar SED models.

Because of this random sampling, and since the probability distributions of the positive- and negative-parity images can differ substantially (with the negative parity images typically showing more pronounced tails towards both low and high magnification; \citealt{Palencia24}), the two images of each star will end up with different brightness (i.e., $m_{\mathrm{AB},j,+}\neq m_{\mathrm{AB},j,-}$). 

Depending on the spatial distribution of stars within the star cluster, and the distance from the star cluster to the caustic in the source plane, the macroimages of the stars can observationally appear concentrated into two well-defined star cluster macroimages (either resolved or unresolved depending on $\mu_t$ and intrinsic star cluster size) of opposite parity in the image plane, or as such images plus separate images formed by one or several stars that -- because of their location in the outskirts of the cluster -- happen to lie closer to the caustic than the rest of the star cluster. The latter images, which can reach extremely high magnification ($\mu\sim 10000$), should appear very close to the critical curve in the image plane and may even blend into a single unresolved image \citep[as has been argued to be the case for the lensed star candidate Earendel at $z\approx 6$;][]{Welch22}. An example of a star cluster at $d_\mathrm{caustic}=10$ pc from the caustic, and the resulting magnification distributions of its stars are shown in Figure~\ref{fig:example_cluster}.

In this paper, we will focus on the appearance of unresolved star cluster images at moderate macromagnification ($\mu_{\mathrm{t}}\sim 10^2$) and will therefore disregard stars in the outskirts of the star cluster that, due to their close proximity to the caustic, end up with $\mu_{\mathrm{t},i}>500$. By summing up the lensed fluxes of all the $\mu_{\mathrm{t},i}\leq 500$ stars in each macroimage of the star cluster, we get an estimate of the expected snapshot brightness and spectral energy distribution of these images. By resampling the $\mu_{+}$ and $\mu_{-}$ distribution for each star within the modelled star cluster many times, we can assess its statistical fluctuations in SED shape. However, since our model does not track the movements of specific stars across the caustic network produced by the microlenses, this machinery does not allow us to predict the expected SED changes between two specific observation epochs. To accomplish the latter, detailed light curve software \citep[e.g.][]{Meena22} should instead be used. 

In this paper, we adopt $B_0=15$ arcsec$^{1/2}$, $d_{i,\mathrm{caustic}}=10$ pc, $\mu_r=1.5$, $z_\mathrm{l}=0.5$ and $z_\mathrm{s}=6$ as our fiducial values, which leads to $\mu_{\mathrm{macro,tot}}\approx 360$ for the two macroimages combined ($\mu_{\mathrm{t}}\approx 120$ and $\mu_\mathrm{r}=1.5$ for each individual macroimage) for a star with the same $x$ coordinate as the star cluster centre ($x=0$). In this case, stars being rejected by the $\mu_{\mathrm{t},i}>500$ criterion (on the basis that the macroimages of such stars may appear separated from that of the star cluster macroimages) would need to be located within $\approx 0.6$ pc from the caustic (at a distance of more than $\approx 9.4$ pc from the cluster center). The image-plane relation between the macromagnification ($\mu_\mathrm{r}\mu_{\mathrm{t},i}$) and the
angular separation $\theta_\mathrm{cc}$ of a macroimage from the critical curve ($\mu_\mathrm{r}\mu_{\mathrm{t},i} = \mu_0 / \theta_\mathrm{cc}$; \citealt{Diego19}) contains a parameter $\mu_0$  that varies between different arcs, but many lens models for well-studied $z\gtrsim 5$ arcs with lensed star clusters or star-forming clumps \citep[e.g.][]{Meena23,Vanzella23,Messa26} would place $\mu_{\mathrm{t},i}>500$ stars (or $\mu_\mathrm{r}\mu_{\mathrm{t},i}$>750, given $\mu_\mathrm{r}=1.5$ in our case) at $<0.1$ arcsec from the critical curve, which would cause a sufficiently bright star in the outskirts of the cluster to be observationally identified as a separate lensed-star candidate source. While the adopted $\mu_{\mathrm{t},i}$ cut-off is somewhat arbitrarily chosen, this will affect only a small fraction of the stars in the very compact star clusters (Section~\ref{subsec:spatial distribution}) that we consider ($\approx 0.1\%$ for our chosen $\mu_{\mathrm{t},i}>500$ limit;  an alternative limit at $\mu_{\mathrm{t},i}>300$ would exclude $\approx 0.3\%$ of the stars.)

For the surface mass density of microlenses, we adopt $\Sigma_\star=10\ M_\odot$ pc$^{-2}$ as our default value. While lower-$z$ lensed stars may be subject to $\Sigma_\star$ a few times higher than this \citep[e.g.][]{Diego24}, macrolensing critical curves move outward from the centre of the cluster lens with increasing redshift. This places them in regions where the intracluster light is typically fainter and the surface mass density stellar microlenses lower, which makes  $10\ M_\odot$ pc$^{-2}$ a suitable choice for the $z\gtrsim 6$ lensing which is the focus of this paper.

Since the M{\_}SMiLe magnification model \citep{Palencia24} does not explicitly take source-size effects into account, some modifications to the M{\_}SMiLe magnification distributions are required in the case of very large sources. M{\_}SMiLe uses a source-plane pixel size of $2\times 10^{-9}$ arcsec, which at our fiducial source redshift of $z_\mathrm{s}=6$ can contain the stellar photosphere for a star of radius up to $\approx 250\ R_\odot$. For evolved stars with larger radii, the probabilities for very large and very low magnifications are expected to be reduced compared to the  M{\_}SMiLe predictions. For example, a 120 $M_\odot$ star in our $Z=0.002$ models can reach a radius of $\approx 1600 R_\odot$ at the end of its lifetime. Even larger stars are possible for our Pop III models, with 300 $M_\odot$ and 2000 $M_\odot$ stars reaching  $\approx 2800\ R_\odot$ and $\approx 5900\ R_\odot$ respectively. To accommodate sources larger than the M{\_}SMiLe pixel size, we adopt the approach of limiting the magnification range for both positive- and negative-parity images to 0.1--3 times the macrolensing magnification of each image ($\mu_\mathrm{t}\mu_\mathrm{r}$) by simply truncating the magnification probability density functions at magnifications beyond these limits. This approach, further discussed in Appendix~\ref{sec:appendix_source_size} is motivated by tests with sources covering up to 24 pixels (matching the $\approx 5900\ R_\odot$ case) in size, and is conservative in the sense that it will systematically underpredict the probability for large variations from the $\mu_\mathrm{t}\mu_\mathrm{r}$ value, due to the sharp truncation. 

M{\_}SMiLe may also underestimate the probability for extremely high magnifications when sources are much smaller than the native resolution ($\ll 250\ R_\odot$ in our case). Taking this into account could slightly increase the variability of smaller compact stars  on the main sequence. Hence, our approach of using the unaltered M{\_}SMiLe probability distributions for these cases ensures that our estimates are on the conservative side.

\subsection{Stellar initial mass function}
\label{subsec:IMF}
For our fiducial models, we assume the stars to obey the \citet{Kroupa01} universal IMF throughout the 0.1--120 $M_\odot$ mass range, i.e. an IMF for high-mass stars similar to that in the local Universe. To consider the possibility that the IMF is modified at very low metallicity and/or high redshift, we also consider an extremely top-heavy IMF with log-flat distribution $\mathrm{d}N/\mathrm{d}M\propto M^{-1}$ \citep[e.g.][]{Magg16} in the case of Pop III stars, with either 2--300 $M_\odot$ or 2--2000 $M_\odot$ as the adopted mass range. 

\subsection{Spatial distribution of stars}
\label{subsec:spatial distribution}
For our modelled star clusters, we adopt the \citet{Elson87} surface brightness profile and assume a constant stellar mass-to-light ratio with radius $r$ (equivalent to assuming no mass segregation and no contribution to the surface brightness from nebular emission). 
Using the \citet{Ryon17} notation for the \citet{Elson87} profile, this leads to a radial distribution function for stars of:
\begin{equation}
P(r)=P(0)\left(1 + \frac{r^2}{a^2} \right)^{-\eta},
\end{equation}
where $a$ is the scale radius in pc and $\eta$ is a somewhat loosely constrained parameter, which we here set to $\eta=1.5$ in rough agreement with many observed young star clusters \citep{Ryon17}.

The scale radius $a$ is related to the effective radius $R_\mathrm{eff}$ in pc through \citep{Brown21}
\begin{equation}
a=\frac{R_\mathrm{eff}}{\sqrt{2^{1/(\eta-1)}-1}}.
\end{equation} 
 
While $R_\mathrm{eff}$ formally represents the circular radius that contains half of the light of the cluster light profile, our assumption on constant mass-to-light radio implies that this also represents the projected half-mass radius of the star cluster.
For young star clusters in the local Universe, $R_\mathrm{eff}$ has been found to depend on the total stellar mass \citep[e.g.][]{Brown21}. However, to accommodate the extremely compact, high-mass ($\sim 10^6 \ M_\odot$, $R_\mathrm{eff}\sim 1$ pc) star clusters recently detected at $z\approx 10$ \citep{Adamo24}, we adopt $R_\mathrm{eff} = 1$ pc as our standard case throughout this paper. 

\begin{figure*}
	\includegraphics[width=\columnwidth]{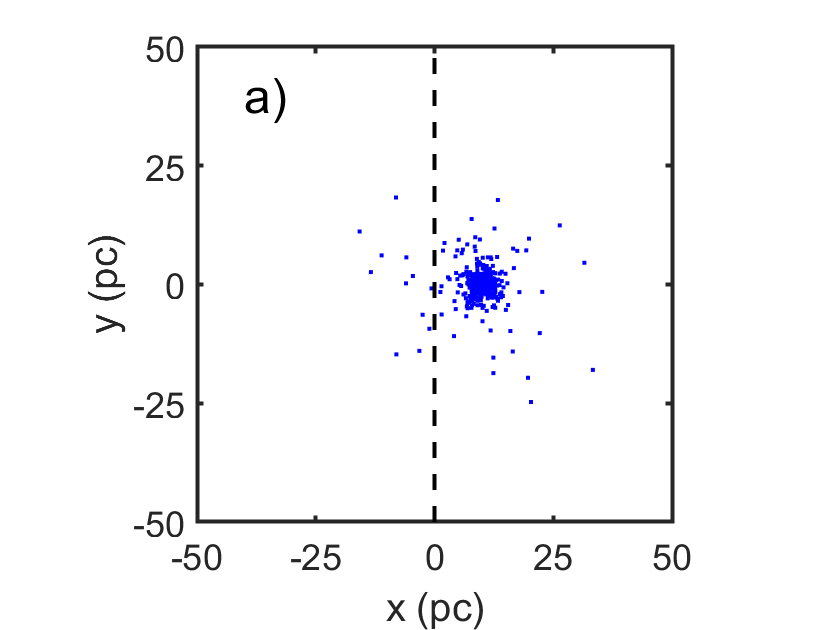}
	\includegraphics[width=\columnwidth]{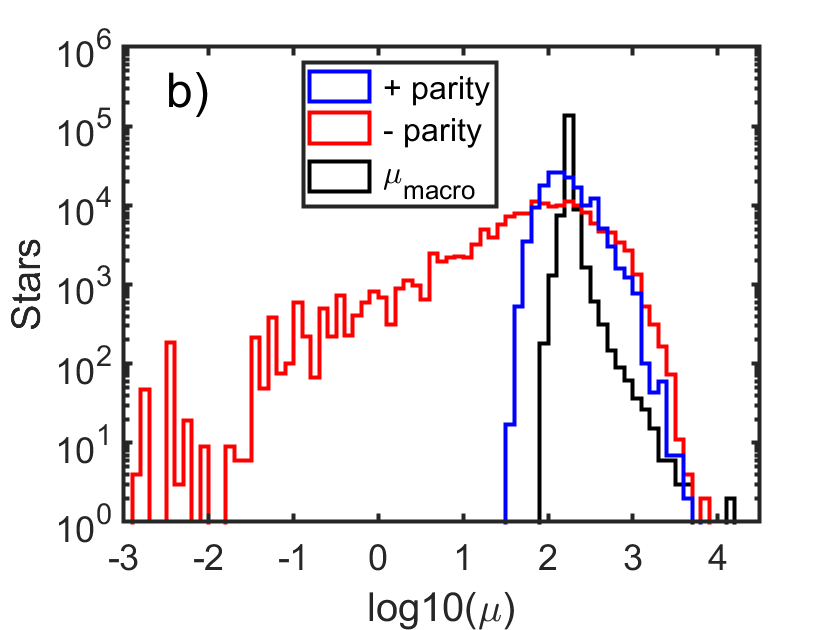}
    \caption{Macro- and microlensing of stars within a $10^5\ M_\odot$ star cluster at a distance of 10 pc from the macrolensing caustic. {\bf a)} Projected spatial distribution of the $\geq 10\ M_\odot$ stars (blue dots) with respect to the caustic (dashed black line).
{\bf b)} Magnification distribution for stars within a single macroimage of the star cluster, when placed at $z=6$, under the assumption of $\Sigma_\star=10\ M_\odot $ pc$^{-2}$ in microlenses in the $z_\mathrm{l}=0.5$ lens plane: positive-parity image (blue line), negative-parity image (red line), macrolensing only (black line, same for positive/negative parity). Stars to the left of the caustic in panel a do not produce counterimages and are not included here.
Due to gravitational microlensing, both the positive- and negative parity distributions are wider than the distribution due to macrolensing only. Compared to the positive-parity distribution, the negative-parity probabiliity distribution exhibits a prominent tail to very low magnifications, but also a subtle boost of the probabilities for very high magnifications.}
    \label{fig:example_cluster}		
\end{figure*}

\subsection{SED models}
\label{subsec:SED models}
In this paper, we use the Muspelheim models \citep{Zackrisson24} for the SEDs of stars as a function of stellar age and Zero Age Main Sequence (ZAMS) mass. For our fiducial SED models, which we consider suitable for typical low-metallicity stars at high redshifts, we adopt the $Z=0.002$ PARSEC v.2.0 stellar evolutionary tracks \citep{Costa25} for single, non-rotating stars in the 2--120 $M_\odot$ mass range and PARSEC v.1.2 \citet{Chen14} tracks for stars at 0.1--1.8 $M_\odot$. These tracks are coupled to the $[M/H]=-1$ stellar atmosphere grid described in \citet{Zackrisson24}. For metal-free Pop III stars, we adopt the \citet{Costa25} PARSEC v.2.0 tracks at $Z=10^{-11}$ for single, non-rotating stars with masses 2--2000 $M_\odot$ and the Pop III stellar atmosphere grid of \citet{Zackrisson24}.
Since these evolutionary tracks include pre-main sequence evolution, and since we assume that all stars within a star cluster start evolving simultaneously from the beginning of their respective track, stars of different mass will reach the ZAMS at slightly different ages (e.g. $\approx 2\times 10^4$, $1\times 10^5$, $4\times 10^5$, $1\times 10^6$ and $2\times 10^6$ yr for Pop III stars with ZAMS masses 300, 50, 10, 5 and 3 $M_\odot$ respectively, and somewhat earlier for the $Z=0.002$ stars). This has an impact on the early SED evolution (at ages up to $\approx 2\times 10^6$ yr) of our modelled star clusters, compared to the case where all stars are assumed to start on the ZAMS at the same time \citep[see also][]{Mitani19}. 

From the stellar spectra produced by Muspelheim as a function of stellar age and ZAMS mass, we form photometric SEDs by computing for the broadband filter AB magnitudes in JWST/NIRcam filters F090W, F115W, F150W, F200W, F277W, F356W and F444W at the source plane redshift $z_\mathrm{s}$ of the stars. However, due to the presence of the Lyman-$\alpha$ break within the F090W filter at our fiducial source redshift ($z_\mathrm{s}=6$), our discussion of colour differences across mirror-image SEDs will focus on a comparison of the F115W and F444W fluxes, as this captures the balance between the short- and long wavelength sides of the observed JWST/NIRCam SED. 

Since the evolutionary tracks used tend to be coarsely sampled at the high-mass end, we interpolate the Muspelheim photometric SED models to a grid with $\Delta M_\mathrm{ZAMS}=0.1\ M_\odot$ before assigning SED models to the stars in our modelled star clusters. By assuming a single age $t$ for all stars within a star cluster, the intrinsic SED of each star in JWST/NIRCam filters is then derived by finding the closest match in Zero Age Main Sequence (ZAMS) mass among our set of interpolated SED models, and then interpolating in age among the available Muspelheim SEDs for that mass. If the age $t$ turns out to be greater than the lifetime of the star (which means that the star has expired), it is not considered further. 

As the phenomenon of mismatched mirror-image SEDs will be more pronounced whenever some part of the star cluster SED is dominated by just a few high-mass stars (with short lifetimes), we throughout this paper limit the discussion to star clusters (and hence stars) with ages up to 40 Myr, which roughly corresponds to the lifetime of $\approx 8\ M_\odot$ ($Z=0.002$) and $\approx 7\ M_\odot$ (Pop III) stars given the adopted stellar evolutionary tracks. 

\section{Mirror images with discrepant SEDs}
\label{sec:normal star cluster}
Mirror images of a lensed star cluster often attain somewhat different brightness \citep[e.g.][]{Adamo24}, due to the presence of local perturbers (e.g. low-mass galaxies, dark matter subhalos in the lens or elsewhere along the line of sight) close to either image. At the same time, it is commonly assumed that the mirror images should exhibit the same SED shape -- this is often used as a criterion when mirror images are observationally identified. Since every star within a star cluster may attain slightly different microlensing magnifications in the two macroimages, we here explore scenarios where bright stars that contribute significantly to the integrated SED are affected by unusually high or low microlensing magnifications, thereby making the SED shapes of the two macroimages differ.

\subsection{Differential magnification within star clusters}
\label{subsec:differential_magnification}
How wide is the distribution of magnifications among the stars in a lensed star cluster expected to be? In Figure~\ref{fig:example_cluster}a, we show an example of the spatial distribution of stars within the source plane for a standard-IMF, $M_\mathrm{tot}=10^5\ M_\odot$ star cluster, under the assumptions outlined in Section~\ref{sec:computational machinery}. Since this star cluster contains $\approx 1.6\times 10^5$ stars, we here plot only the positions of stars of mass $\geq 10\ M_\odot$ ($\approx 900$ stars) to avoid cluttering. Despite assuming a very compact star cluster ($R_\mathrm{eff}=1$ pc), small numbers of stars are found out to a radius of $\approx 25$ pc, because of the adopted surface mass distribution, and a few stars may lie sufficiently close to the caustic that they could be detected as individual lensed stars separate from the unresolved macroimages of the star clusters produced by this configuration. 

In Figure~\ref{fig:example_cluster}b, we show the corresponding magnification distribution (due to a combination of macro- and microlensing) of all $\approx 1.6\times 10^5$ stars within the positive- and negative-parity macroimages, compared to the corresponding macrolensing-only magnification ($\mu_\mathrm{t}\mu_\mathrm{r}$) within a single image. Here, we see that the magnification distribution of the negative-parity image is wider than that of the positive-parity image \citep[an effect discussed in][]{Palencia24}, with a prominent tail to very low magnifications, and that both magnification distributions are wider than that produced by macrolensing alone. As Figure~\ref{fig:example_cluster}b is meant to represent a generic case, all stars born within the cluster and falling to the right of the caustic in Figure~\ref{fig:example_cluster}a are included. Hence, no age effects, source-size effects, nor the $\mu_\mathrm{t}\leq 500$ cut-off described in Section~\ref{subsec:lensing}, have been applied. However, source-size effects would only apply to a minor and age-dependent fraction of these stars, and only $\sim 0.1\%$ of the stars are at $\mu_\mathrm{t}> 500$ in this figure.

\subsection{The impact of differential magnification on the integrated SED of macroimages}
\label{subsec:differential_magnification_example}
How would the distribution of magnifications among the member stars (Figure~\ref{fig:example_cluster}b) affect the SEDs of the entire star cluster macroimages? In a young star cluster ($\lesssim 40$ Myr) with a standard IMF, most high-mass stars will be found close to their ZAMS position at high $T_\mathrm{eff}$, and only a few (sometimes just one, or none at all, depending on $M_\mathrm{tot}$, the IMF sampling and the exact age of the star cluster) of the most massive ones will be in a more evolved state at a significantly lower $T_\mathrm{eff}$ (either because the main sequence extends to low $T_\mathrm{eff}$ for high-mass, low-metallicity stars, or because stars reach these stages as part of their post-main sequence evolution; e.g. \citealt{Costa25}). While the high-$T_\mathrm{eff}$ stars dominate the rest-frame UV SED of the star cluster, the low-$T_\mathrm{eff}$ evolved stars, if present, may dominate the SED in the rest-frame optical (and near-infrared, although this part is only sampled by JWST/NIRCam for star clusters at $z\lesssim 4$).

\begin{figure*}
	\includegraphics[width=\columnwidth]{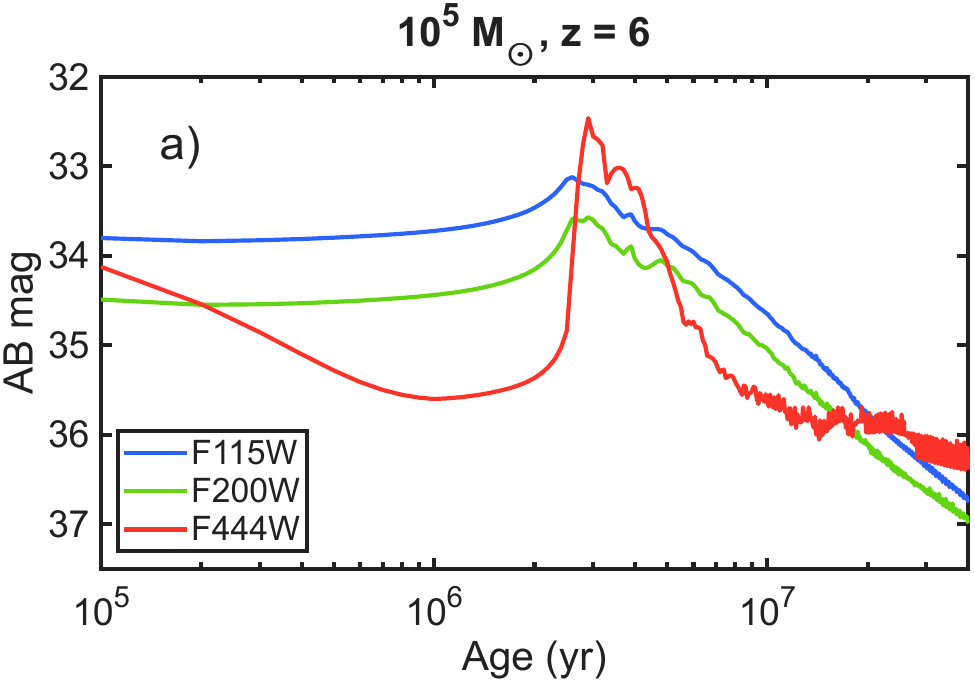}
	\includegraphics[width=\columnwidth]{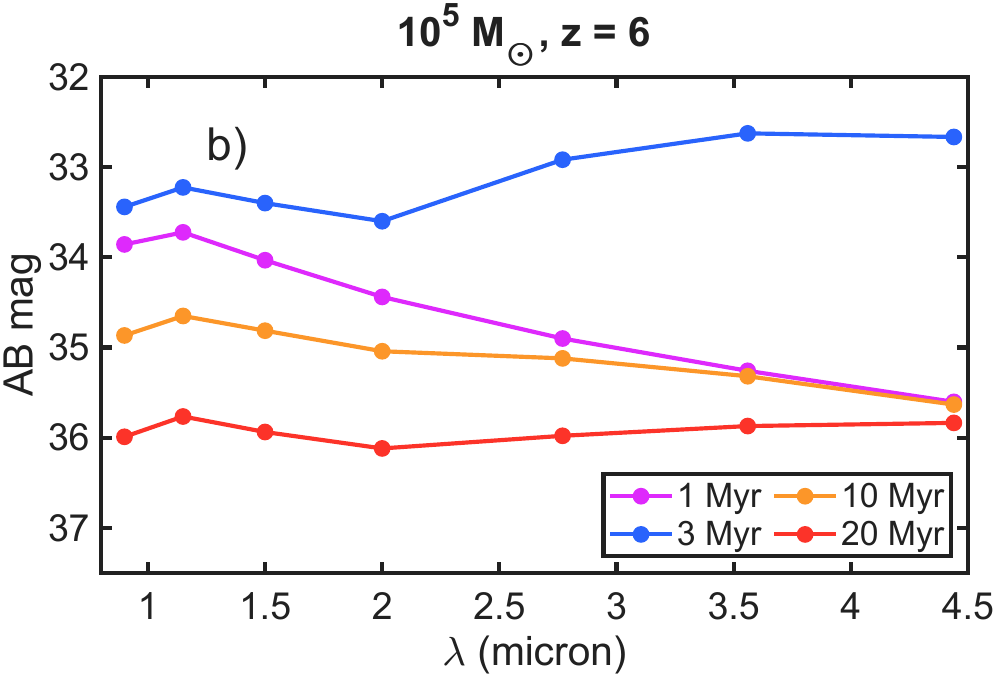}
    \caption{The photometric evolution in JWST/NIRCam bands of the star light from a young, unlensed $10^5\ M_\odot$, $Z=0.002$ single-age star cluster at $z=6$, under the assumption of a fully-sampled \citet{Kroupa01} IMF from 0.1--120 $M_\odot$: {\bf a)} Brightness evolution in filters F115W, F200W and F444W as a function of age; {\bf b)} JWST/NIRcam SEDs at ages 1, 3, 10 and 20 Myr. The contribution from evolved, red stars is seen as an upturn in the SEDs for wavelengths above 2 $\mu$m at ages 3 Myr and above.}
    \label{fig:integrated_population}		
\end{figure*}

Let's consider two different cases: one where all the massive stars within the star cluster are at high $T_\mathrm{eff}$ ($\gtrsim 30000$ K), and one where a few massive stars have evolved to low $T_\mathrm{eff}$ ($\lesssim 10000$ K). In appendix~\ref{sec:appendix_light_contribution}, we show the relative light contribution from stars of different ZAMS mass to the overall star cluster JWST/NIRCam wide-band fluxes, in examples of a young star cluster at $z=6$ where these two cases apply. In the first case, the distribution of magnifications among the individual stars should have very little impact on the relative shape of the SED of the macroimages, since all these high-$T_\mathrm{eff}$ stars have fairly similar SED shapes throughout the rest-frame UV and optical \citep[e.g.][]{Zackrisson24}. In the second case, the situation is different, since attaching a high or low magnification to one of the few low-$T_\mathrm{eff}$ stars will upset the balance between the short- and long-wavelength parts of the rest-frame UV/optical SED. Altering the magnifications of some of the brightest high-$T_\mathrm{eff}$ stars can also upset this balance, albeit to a much smaller extent, since the short-wavelength side of the SED carries significant contributions from a large number of stars close to their ZAMS temperatures, whereas the long-wavelength side carries significant contributions from just a few evolved stars (at least in the case of a standard IMF).

\subsection{Observable SED differences between mirror images}
\label{subsec:observable SED differences}
\begin{figure*}
	\includegraphics[width=\columnwidth]{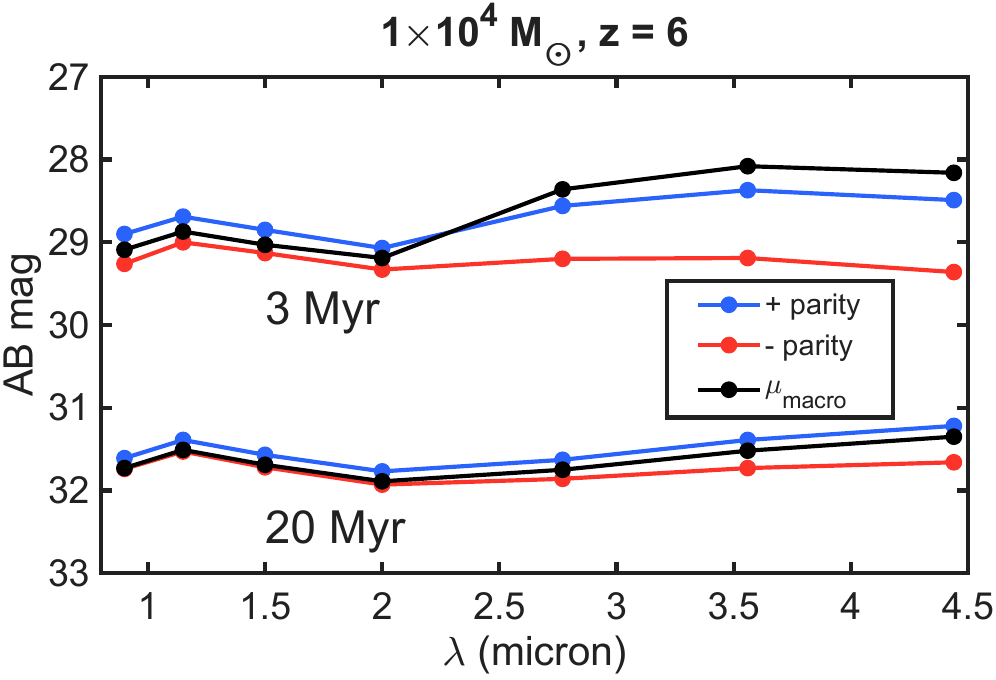}
	\includegraphics[width=\columnwidth]{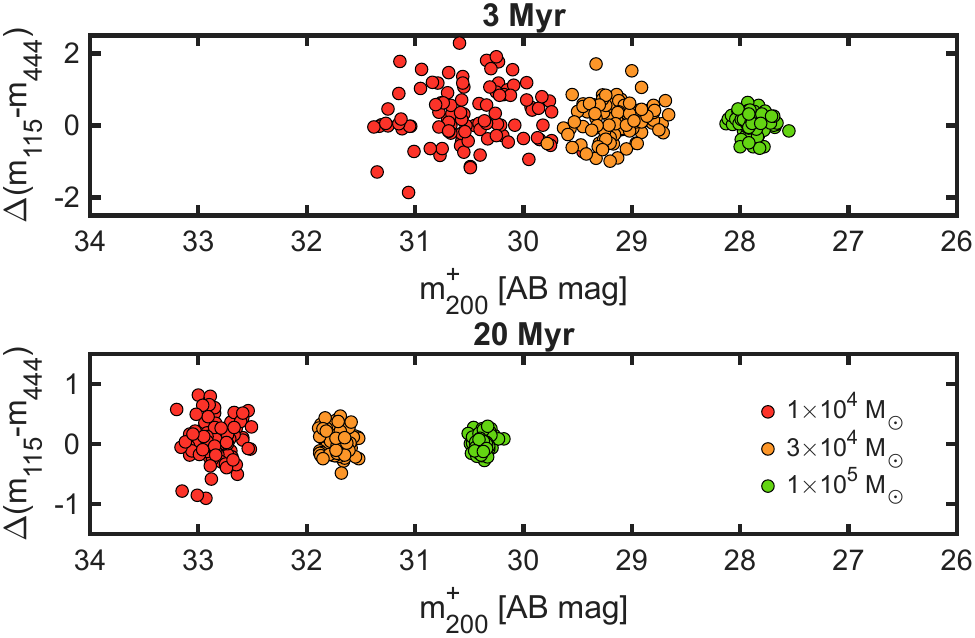}
    \caption{Discrepant SEDs of star cluster mirror images at $z=6$, formed by  $Z=0.002$ star clusters at 10 pc from the macrocaustic and affected by $\Sigma_\star=10\ M_\odot$ pc$^{-2}$ microlensing.	{\bf Left)} Examples of JWST/NIRCam SEDs of positive (blue line) and negative-parity (red line) mirror images of a $3\times 10^4\ M_\odot$ star cluster at ages 3 and 20 Myr. The black line shows the corresponding SEDs predicted by macrolensing alone (the same for both mirror images). {\bf Right)} Variation in $m_{115}-m_{444}$ colour between positive- and negative-parity images ($\Delta(m_{115}-m_{444})=(m_{115}-m_{444})^{+} - (m_{115}-m_{444})^{-}$), as a function of the F200W apparent magnitude of the positive-parity image ($m_{200}^{+}$), for star clusters with total masses $1\times 10^4$ (red circles), $3\times 10^4$ (orange circles), $1\times 10^5\ M_\odot$ (green circles) at ages 3 and 20 Myr. Every differently coloured cloud of points correspond to 100 Monte Carlo realizations. The colour differences between the mirror images are seen to be larger at 3 Myr than at 20 Myr (please note the different y-axis limits in the right panel). At fixed age, the colour differences between the mirror images is also seen to become larger for lower-mass star clusters, while at the same time occurring at fainter overall SED magnitudes ($m_{200}^{+}$).}
\label{fig:discrepant_SEDs}		
\end{figure*}
Under what conditions are these SED discrepancies between mirror images expected to be sufficiently prominent to be detected in typical JWST observations of cluster-lens fields?

To explore over what age ranges such effects are likely to occur, we in Figure~\ref{fig:integrated_population}a show the predicted intrinsic brightness evolution at $z=6$ in filters JWST/NIRCam F115W, F200W and F444W for a low-metallicity, $M_\mathrm{tot}=10^5\ M_\odot$, instantaneous-burst (i.e., single-age) stellar population up to an age of 40 Myr, under the assumption of a fully-sampled \citet{Kroupa01} IMF. All three filter fluxes display a peak around 2--3 Myr, when the most massive stars have evolved off the main sequence and contribute maximally to the rest-frame ultraviolet and optical wavelengths covered by these filters, followed by a decline at higher ages as more and more of the massive stars reach the end of their lifetimes. The evolution of the F444W flux is, however, more dramatic and somewhat different than in the F115W and F200W filters, with a decline for the first Myr due to pre-main sequence evolution, followed by a sharp rise above the F115W and F200W fluxes at $\approx 3-5$ Myr, followed by a drop down to a more slowly-evolving plateau at 10-40 Myr. This behaviour at $\gtrsim 3$ Myr is due to the appearance of yellow and red supergiants at $T_\mathrm{eff}\lesssim 10^4$ K, formed from $M\gtrsim 40\ M_\odot$ stars in these models. 

The corresponding evolution in the full JWST/NIRCam SED at ages 1, 3, 10 and 20 Myr is shown in Figure~\ref{fig:integrated_population}b, where the contribution from these bright, lower-$T_\mathrm{eff}$ stars is seen in filters longward of 2$\mu$m at ages 3, 10 and 20 Myr.
Due to the brief duration of these red, bright phases ($\sim 10^5$ yr), the star cluster will at any given time contain only a small number of such stars. Based on the adopted models, the total number of stars in a 3-Myr old $10^5\ M_\odot$ star cluster that simultaneously display $T_\mathrm{eff}\leq 10^4$ K and bolometric luminosities $\geq 10^5\ L_\odot$ is $\approx 10$ under the assumption of a fully-sampled IMF. As the most massive stars die off, the F444W flux at ages of 10--40 Myr become superseded by a larger number (a few tens) of somewhat less-luminous warm/cool supergiants with $T_\mathrm{eff}\leq 10^4$ K and $10^{4-5}\ L_\odot$.   

In such cases, where the long-wavelength part of the star cluster SED is dominated by a limited number of bright, red stars, and gravitational microlensing results in significantly different positive- and negative-parity magnifications for some of these stars, then mirror images with discrepant SEDs may result. 

The discrepant-SED effect is expected to become smaller for high-mass star clusters due to more red stars co-existing at the same time, which dampens these fluctuations, and larger for low-mass star clusters. However, at sufficiently low star cluster masses ($\lesssim 10^4\ M_\odot$), some young systems may not contain any red/yellow supergiants at all due to IMF sampling effects. Even if they do, the overall brightness of the star cluster in the short-wavelength part of the SED may be too low for detection with JWST for macromagnifications $\sim 100$. This means that there will be a limited range of star cluster masses over which the discrepant-SED phenomenon is observable, given the detection limits of JWST observations of cluster-lens fields. 

\subsection{Mass and age dependence of mirror-image SED discrepancies}
\label{subseq:mass and age dependence}
Figure~\ref{fig:discrepant_SEDs} illustrates the age and mass trends suggested by the discussion in Section~\ref{subsec:observable SED differences}. In Figure~\ref{fig:discrepant_SEDs}a, we show example SEDs of positive- and negative-parity mirror images of a $3\times 10^4\ M_\odot$ star cluster at $z=6$, located at 10 pc from the macrocaustic, under the assumption of star cluster ages 3 or 20 Myr. For these modelled star clusters, the IMF has been randomly sampled, which introduces small differences between the intrinsic SED and those depicted in Figure~\ref{fig:integrated_population} (which assume complete sampling of the IMF). Due to microlensing, the 3 Myr-old cluster also shows pronounced differences between the positive- and negative-parity mirror images (with the F115W-F444W color differing by $\approx 0.6$ mag) at an overall SED brightness in the $\approx $ 28.5--29.5 AB mag range. Moreover, the negative-parity image deviates substantially from the SED shape expected from macrolensing alone, with a less pronounced red bump at observed wavelengths $>2\ \mu$m. In the 20 Myr example, the difference is smaller (with the F115W-F444W color differing by $\approx 0.3$ mag between the images) due to the long-wavelength side of the SED being dominated by a larger number of red stars. The observed brightness of this star cluster is furthermore just 31--32 mag, which places it well beyond the detection limit of typical cluster-lens observations with JWST. 

To illustrate the $M_\mathrm{tot}$ dependence, we in Figure~\ref{fig:discrepant_SEDs}b show the scatter in the difference $\Delta(m_{115}-m_{444})$ between the F115W-F444W colours of the positive- and negative-parity mirror images ($\Delta(m_{115}-m_{444})=(m_{115}-m_{444})^{+} - (m_{115}-m_{444})^{-}$) as a function of the F200W AB magnitude in the positive-parity image ($m_{200}^{+}$), for 100 realizations of lensed star clusters (10 intrinsically separate clusters with different stellar masses and positions of stars, each subject to 10 reshuffles of the magnifications for every star) with $M_\mathrm{tot}=1\times 10^4$, $3\times 10^4$, $1\times 10^5\ M_\odot$ at ages 3 and 20 Myr. Here, we see that at fixed star cluster age, the $\Delta(m_{115}-m_{444})$ variation increases with decreasing star cluster mass, in agreement with the expectations from the previous discussion. However, at 20 Myr, the overall brightness of the SED (here quantified by $m_{200}^{+}$) is too low ($\lesssim 30$ AB mag) for such star clusters to be studied with JWST in typical cluster-lens surveys (with photometric 5$\sigma$ limits in the range 28--29 AB mag), and can only be probed in the very deepest surveys (5$\sigma$ limits of 30--31 AB mag, which would be sufficient to detect our $10^5\ M_\odot$ model star clusters).

In the lower-mass star clusters ($\sim 10^4\ M_\odot$) of Figure~\ref{fig:discrepant_SEDs}b, IMF sampling effects and microlensing combine to shape the overall $m_{200}^{+}$ and  $\Delta(m_{115}-m_{444})$ distributions seen. IMF sampling gives rise to an intrinsic spread in the unlensed $m_{200}$ flux among star clusters, and -- while giving the same flux boost/depletion to the two components of a specific mirror-image pair -- broadens the lensed $m_{200}^{+}$ (and $m_{200}^{-}$) distribution across a population of star cluster images. The effect of IMF sampling on $\Delta(m_{115}-m_{444})$ is more complex, and alters the overall distribution by simultaneously increasing the fraction of $\Delta(m_{115}-m_{444})\approx 0$ cases (caused by star clusters without any bright, red stars) and the number of high-$\Delta(m_{115}-m_{444})$ outliers in the tail of the distribution.

\subsection{Masses and ages of star clusters for which mismatched mirror-image SED shapes may be detected with JWST}
\begin{figure}
	\includegraphics[width=\columnwidth]{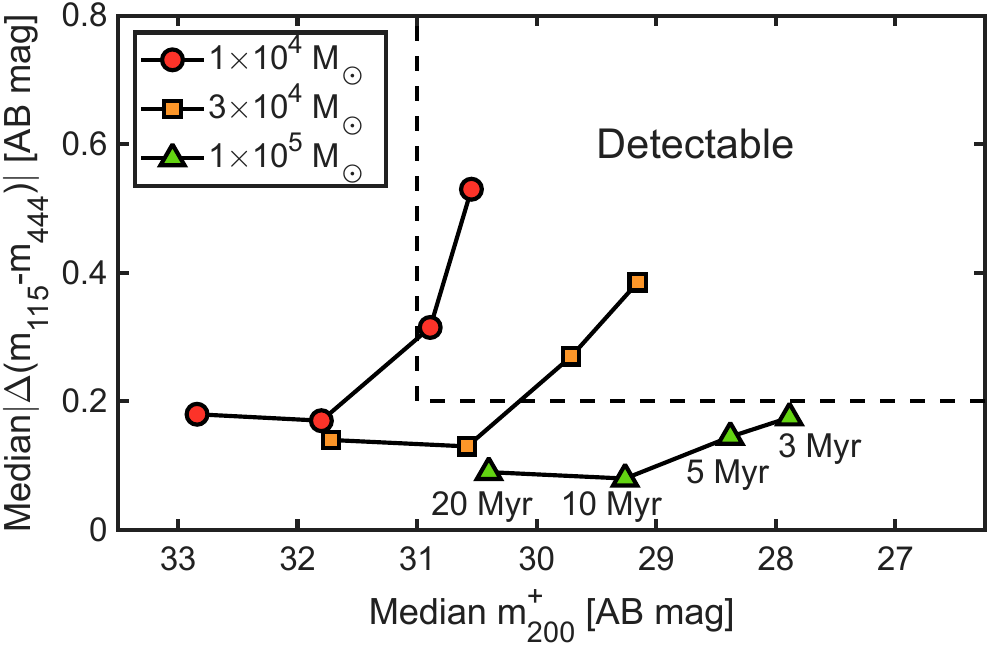}
    \caption{Median absolute colour difference between the F115W and F444W filters for mirror images of young $Z=0.002$, $z=6$ star clusters affected by microlensing with $\Sigma_\star=10\ M_\odot$ pc$^{-2}$, as a function of the median apparent F200W flux of the positive-parity star cluster image. Models for star clusters of mass $1\times 10^4\ M_\odot$ (red circles), $3\times 10^4\ M_\odot$ (orange squares) and $1\times 10^5\ M_\odot$ (green triangles) are shown at ages 3, 5, 10 and 20 Myr (from right to left). The dashed line indicates the part of the diagram where we consider that colour differences may be detectable in JWST observations of cluster-lens fields (see main text for details). For star clusters in the range $1\times 10^4$--$3\times 10^4\ M_\odot$, the median colour difference is in the observable range for ages $\approx 3$--5 Myr.}
\label{fig:average coldiff}		
\end{figure}

To illustrate the amplitude of typical colour differences between star cluster macroimages, we in Figure~\ref{fig:average coldiff} show the median absolute $m_{115}-m_{444}$ colour difference between the positive- and negative-parity images ($|\Delta(m_{115}-m_{444})|=|(m_{115}-m_{444})^{+} - (m_{115}-m_{444})^{-}|$) of an ensemble of simulated star cluster mirror-image pairs, as a function of the median apparent F200W magnitude of the positive-parity images among these pairs. Results are shown for $z=6$ star clusters at ages 3--20 Myr, in the case of $M_\mathrm{tot}=1\times 10^4$, $3\times 10^4$ and $1\times 10^5\ M_\odot$. For all three sets of star cluster masses, the median $|\Delta(m_{115}-m_{444})|$ colour difference is seen to drop from 3 Myr to 10 Myr, and then remain constant or slightly increase again at 20 Myr. This happens because of a decrease from 3 to 10 Myr in the number of red, very bright stars that drive the discrepant SED-phenomenon, followed by an increase in the number of such stars at 20 Myr. This evolutionary behaviour is also reflected in the overall $m_{115}-m_{444}$ colour evolution of $Z=0.002$ star clusters at these ages (Figure~\ref{fig:integrated_population}), which is redder at 3, 5 and 20 Myr than at 10 Myr.

In Figure~\ref{fig:average coldiff}, we also indicate the region where we consider discrepant SEDs to be detectable given observational limitations: $m_{200}<31$ AB mag due to the depth of the deepest cluster-lens observations with JWST (so far, the GLIMPSE survey has provided the deepest imaging, reaching $\approx 30.8$ AB mag at $5\sigma$ in F200W; \citealt{Atek25}) and $|\Delta(m_{115}-m_{444})|>0.2$ mag, which requires errors of $\leq 0.14$ mag per filter. While it could be argued that photometry on unresolved objects several magnitudes above the survey detection limit should be detectable with errors smaller than this, the error on the photometry of lensed star clusters is usually not dominated by the signal-to-noise per pixel, but by the difficulties in photometrically separating star cluster images from adjacent features in the arc (usually other star clusters), from the diffuse light of the arc itself and from the intracluster light. The quoted errors of star cluster images in arcs are usually no better than $\approx 0.1$--0.2 mag per filter, with limited dependence on their actual brightness \citep[e.g.][]{Adamo24}. Figure~\ref{fig:average coldiff} suggests that discrepant mirror-image SEDs may be detectable in $1\times 10^4\ M_\odot$ and  $3\times 10^4\ M_\odot$ star clusters at ages 3--5 Myr, but typically not at substantially higher cluster masses or ages. Since the $1\times 10^5\ M_\odot$ model displays a median $|\Delta(m_{115}-m_{444})|$ that lies just below the detectable range at 3 Myr, some clusters of this type may occasionally exhibit detectable SED differences between their mirror images (given the scatter in $\Delta(m_{115}-m_{444}$ evident from Figure~\ref{fig:discrepant_SEDs}b), but most will not. 
 
As discussed in Section~\ref{subseq:mass and age dependence}, 
IMF sampling effects become important for the distribution of mirror-image colour differences across a population of lensed star clusters when the mass of the star clusters approach $\lesssim 10^4\ M_\odot$. Numerical tests carried out by comparing the results from simulated 3 Myr-old, $10^4\ M_\odot$ star clusters with randomly IMF-sampled stellar masses to one with a fixed, typical distribution of stellar masses indicates that the net effect of IMF sampling is to slightly reduce the median colour difference ($|\Delta(m_{115}-m_{444})|$) between mirror images, since IMF sampling increases the fraction of $|\Delta(m_{115}-m_{444})|\approx 0$ objects (caused by star clusters lacking bright red stars) in the simulated population.

\section{Discrepant SEDs in the case of top-heavy IMFs}
\label{sec:extreme IMF}
What would happen to the effects outlined in
Section~\ref{sec:normal star cluster} if the star cluster IMF were more top-heavy than our standard case (\citealt{Kroupa01} for masses 0.1--$120 M_\odot$)? If one considers a star cluster with fixed $M_\mathrm{tot}$, keeps the stellar mass range intact, but increases the fraction of high-mass stars by changing the power-law slope of the upper IMF, then this would result in a greater number of massive stars overall. An IMF modification of this type would also boost the number of red, evolved stars at star cluster ages where these may form, as well as boosting the overall brightness of the star cluster. In combination, this change in power-law slope at fixed $M_\mathrm{tot}$ would {\it reduce} the tendency for this star cluster to display mirror-images with discrepant SEDs when lensed, since this effect relies on having the intrinsic SED at the long-wavelength end dominated by a {\it small} number of bright red stars, not a large one. At fixed intrinsic brightness (instead of fixed $M_\mathrm{tot}$), star clusters with \citet{Kroupa01} 0.1--$120 M_\odot$ IMFs and flatter (more top-heavy) power-law IMF slopes throughout the same stellar mass range would result in approximately the same number of such stars. Hence, any modifications of this type are likely to have very small effects on $|\Delta(m_{115}-m_{444})|$ at fixed apparent brightness (F200W$^{+}$). This argument is similar to that presented by \citet{Dai21} for why the temporal variation of star cluster images is only weakly affected when switching to their top-heavy IMF case.

However, the situation changes if one extends the mass range of stars in our models to $\gg 100 M_\odot$, and moreover adopts an IMF that implies that a non-negligible fraction of all stars would attain such high masses. In this case, star clusters can become intrinsically very bright during the lifespans of the most massive stars, and also get lensed into the detectable regime for much lower $M_\mathrm{tot}$, with both the short-wavelength and long-wavelength parts of the SED dominated by a very small number of extremely bright stars. In this scenario, SED variations between mirror images can potentially grow stronger than in the standard-IMF case. 

\begin{figure}
	\includegraphics[width=\columnwidth]{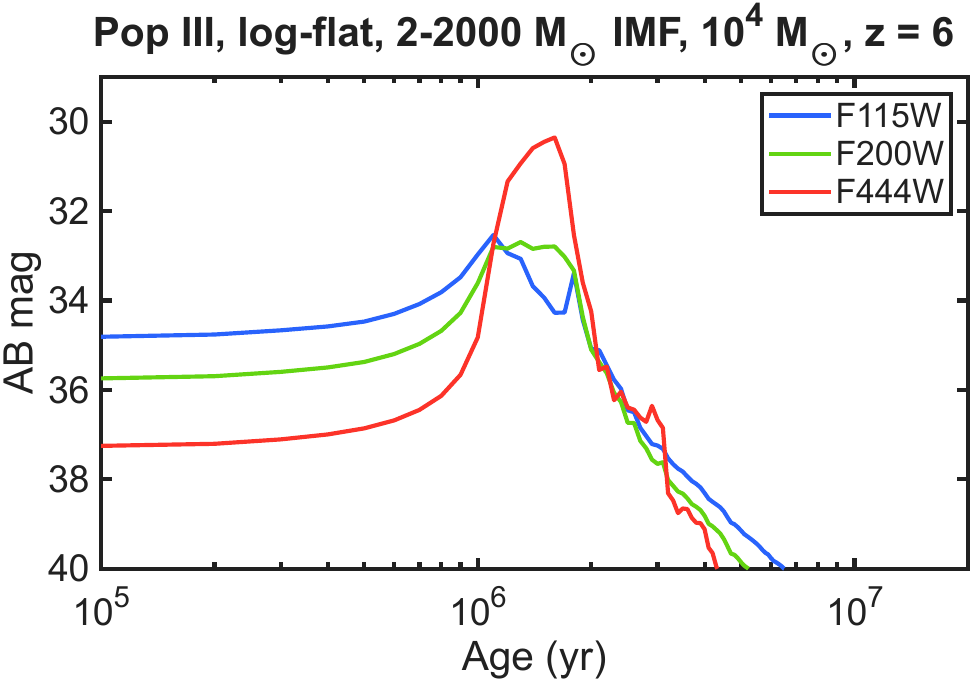}
    \caption{Photometric evolution in JWST/NIRCam filters F115W, F200W and F444W of the star light from a young, unlensed $10^4\ M_\odot$, Pop III single-age star cluster at $z=6$, under the assumption of a fully-sampled log-flat IMF from 2--2000 $M_\odot$. At ages $\approx 1$--2 Myr, the star cluster becomes very bright in the F444W band as stars in the 300--2000 $M_\odot$ mass range reach $T_\mathrm{eff}<2\times 10^4$ K states at the end of their lives.}
\label{fig:PopIII_1e4_colevol}		
\end{figure}

In Figure~\ref{fig:PopIII_1e4_colevol}, we show the intrinsic photometric evolution of a $10^4\ M_\odot$ Pop III star cluster at $z=6$ with a log-flat IMF throughout the 2--2000 $M_\odot$ range. Here, we assume the IMF to be fully sampled to illustrate the typical evolution, although such a star cluster would be subject to substantial IMF sampling effects, as the average stellar mass is $\approx 300\ M_\odot$ and the number of stars just $\approx 40$. At ages 1--2 Myr, this $10^4\ M_\odot$ Pop III star cluster attains a brightness in the F115W and F200W filters comparable to the peak brightness of the $10^5\ M_\odot$ (i.e., an order of magnitude more massive), $Z=0.002$ standard-IMF star cluster depicted in Figure~\ref{fig:integrated_population}, yet a F444W flux up to more than 2 mag brighter, due to the 300--2000 $M_\odot$ stars in the \citet{Costa25} set of stellar evolutionary tracks all reaching their extremely bright $T_\mathrm{eff}<2\times 10^4$ K end stages in this age range.

\begin{figure*}
	\includegraphics[width=\columnwidth]{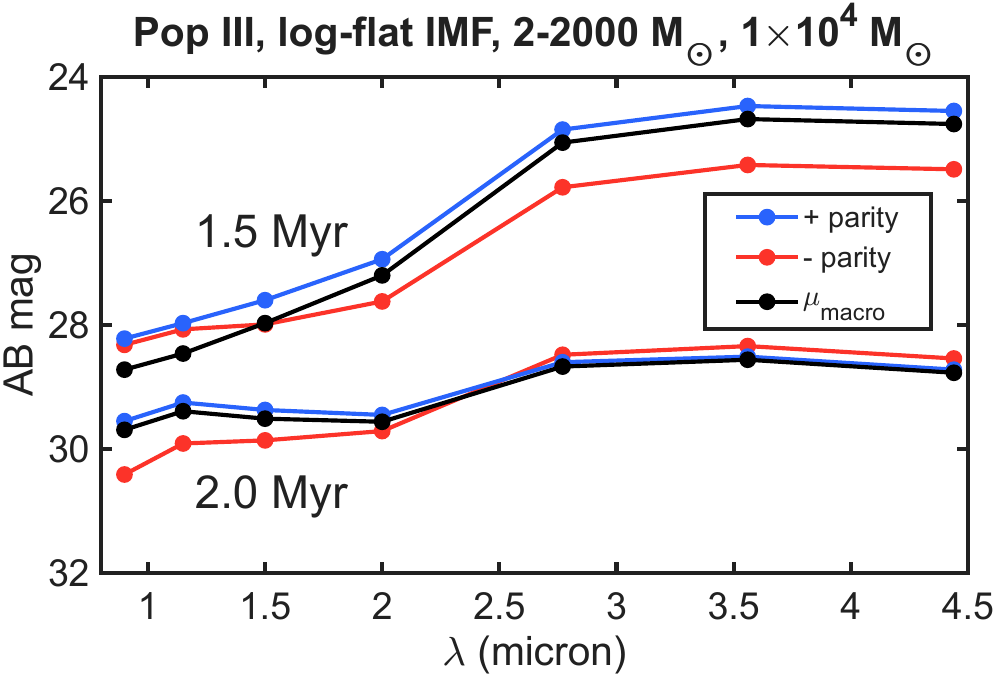}
	\includegraphics[width=\columnwidth]{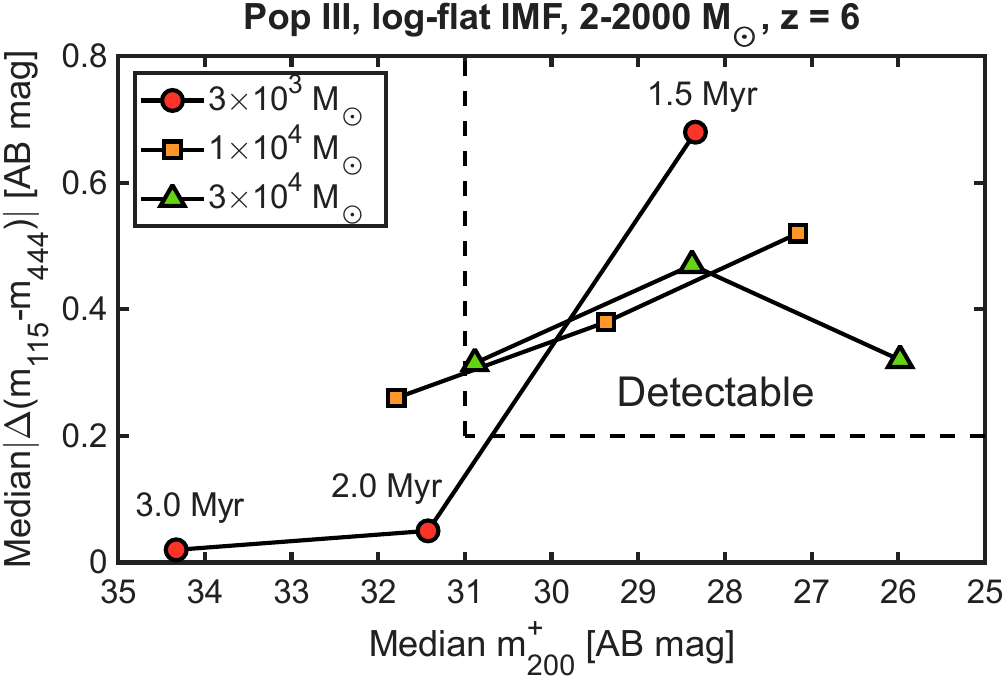}
  \caption{Discrepant mirror image SEDs for $z=6$ Pop III star clusters with a log-flat IMF throughout the 2--2000 $M_\odot$ mass range. {\bf Left:} Examples of JWST/NIRCam SEDs of positive (blue line) and negative-parity (red line) mirror images of a $1\times 10^4\ M_\odot$ star cluster at ages 1.5 and 2.0 Myr. The black line shows the corresponding SEDs predicted by macrolensing alone (the same for both mirror images). {\bf Right:} Median absolute colour difference between the F115W and F444W filters for mirror images as a function of the median apparent F200W flux of the positive-parity star cluster image. Models for star clusters of mass $3\times 10^3\ M_\odot$ (red circles), $1\times 10^4\ M_\odot$ (orange squares) and $3\times 10^4\ M_\odot$ (green triangles) are shown at ages 1.5, 2.0 and 3.0 Myr (from right to left), in relation to the part of the diagram (dashed line) where we consider colour differences to be detectable. For  $3\times 10^4\ M_\odot$ star clusters, discrepant SEDs in the detectable regime are produced at 1.5--3.0 Myr, whereas lower-mass star clusters ( $1\times 10^4\ M_\odot$ and $3\times 10^3\ M_\odot$) only produce discrepant SEDs during a fraction of this age span. At median $m_{200}^{+}<29$ AB mag, the typical colour differences between mirror images in the detectable regime are larger than the corresponding predictions for a standard IMF (Figure~\ref{fig:average coldiff}). The behaviour of the $3\times 10^3\ M_\odot$ models at 2--3 Myr (very faint, with small SED difference between mirror images) is caused by IMF sampling effects due to the small number of stars present within such star clusters, and many cases without bright red stars.}
\label{fig:PopIII_example_and_average_coldiff}		
\end{figure*}

In the left-hand panel of Figure~\ref{fig:PopIII_example_and_average_coldiff}, we show examples of the positive- and negative-parity image SEDs of such Pop III star clusters at ages 1.5 Myr (close to peak brightness in F444W) and 2.0 Myr (more than 3 mag fainter than peak brightness in F444W), but otherwise using the same assumptions as in Section~\ref{sec:normal star cluster}. Not only is the overall discrepancy between the SEDs of the macroimages large (
$|\Delta(m_{115}-m_{444})|\approx 0.8$ mag in both cases), but the difference in SED shape is moreover larger on the short-wavelength side of the SED than on the long-wavelength side (unlike the cases shown for the standard-IMF situation in Figure~\ref{fig:discrepant_SEDs}a). This happens because both sides of the SED are now dominated by a small number of very bright stars, and since the stars dominating the long-wavelength part have very large radii, their probability of attaining very high microlensing magnifications is lower. 

In the right-hand panel of Figure~\ref{fig:PopIII_example_and_average_coldiff}, we show the  
median absolute $m_{115}-m_{444}$ colour difference between the positive- and negative-parity images, as a function of the median apparent F200W magnitude of the positive-parity image, for these Pop III star clusters at total masses of $3\times 10^3$, $1\times 10^4$ and $3\times 10^4\ M_\odot$ and ages 1.5, 2.0 and 3.0 Myr. At $m_{200}^{+}<29$ AB mag, these Pop III star clusters exhibit larger median colour differences than the standard-IMF star cluster shown in Figure~\ref{fig:average coldiff}. In the case of the $3\times 10^3 M_\odot$ star cluster, both the brightness and median colour difference drops dramatically at ages above 1.5 Myr, because just a handful of stars survive to higher ages in such clusters, which means that many such objects will lack the very brightest stars (giving faint F200W) and/or stars with sufficiently different $T_\mathrm{eff}$ to allow the short-wavelength and long-wavelength parts of the SED to be dominated by different stars (giving low $m_{115}-m_{444}$ colour difference between the macroimages). 

In Appendix~\ref{sec:appendix_PopIII_2_300_logflat}, we show the corresponding predictions for Pop III star cluster with an equally extreme (log-flat) IMF slope, but with an upper mass limit of 300 $M_\odot$ instead of 2000 $M_\odot$. As expected from the discussion above, this results in less extreme   
$m_{115}-m_{444}$ colour differences between mirror images than when the IMF is extended to 2000 $M_\odot$, but can still give rise to detectable effects at $m_{200}^{+}<29$ AB mag.

\section{Discussion}
\label{sec:discussion}
In previous sections, we have argued that microlensing in certain situations may allow mirror images of strongly lensed star clusters to appear with discrepant SED shapes. Here, we discuss various other effects that are likely to affect the manifestation of this phenomenon and comment on some shortcomings in the computational machinery currently used to predict the size of the effect. Finally, we comment on the practicalities of using colour differences versus single-filter variability in mirror images as a search for star clusters with extremely top-heavy IMFs.  

\subsection{High-luminosity warm/cool supergiants at low metallicites}
The Humphreys-Davidson limit \citep{Humphreys79} represents an empirical luminosity ceiling of  $L_\mathrm{bol}\sim 10^{5.6} L_\odot$, above which warm/cool supergiants ($T_\mathrm{eff}<10000$ K)  become very scarce in observational samples. While previously established to hold for the Milky Way, Large Magellanic Cloud and Small Magellanic Cloud \citep[e.g.][]{Davies18}, and thereby suggesting an upper mass limit of $M_\mathrm{ZAMS}\approx 30$--$40\ M_\odot$ for stars at $Z\gtrsim 1/5 Z_\odot$ that typically evolve into these evolutionary stages, several sets of stellar evolutionary tracks \citep{Szecsi20,Costa25} predict that more luminous warm/cool supergiants, with higher progenitor masses, should be able to form at lower metallicities. However, \citet{Schootemeijer25} recently presented observational data that suggest the Humphreys-Davidson limit to extend down to the metallicity of IZw18 ($\sim 1/40\ Z_\odot$), seemingly in conflict with these models. The metallicity dependence of the mass/luminosity limit for stars to evolve into warm/cool supergiants clearly has an impact on the effects explored in the current paper. If low-metallicity stars with $M_\mathrm{ZAMS}\gtrsim 40\ M_\odot$ rarely evolve to $T_\mathrm{eff}<10000$ K in the lensed star clusters detected at $z\gtrsim 6$, then discrepant mirror-image SEDs are unlikely to emerge until these star clusters have reached ages of $\geq 5$ Myr, at which point only very small colour differences between the mirror images are expected 
(Figure~\ref{fig:average coldiff}). 

In Section~\ref{sec:extreme IMF}, we argued that Pop III star clusters with very massive stars and a top-heavy IMF at certain ages could attain larger colour differences $|\Delta(m_{115}-m_{444})|$ between their mirror images than low-metallicity ($Z\approx 0.1\ Z_\odot$) star clusters subject to a standard IMF. This assumes that the Humphreys-Davidson limit is violated in the Pop III regime, i.e. that Pop III stars at $M\gg 100\ M_\odot$ stars evolve to become warm/cool supergiants with $L_\mathrm{bol}\gg 10^{5.6}\ L_\odot$  at the end of their lifetimes (and that this phase is not too brief). While this is predicted by several independent sets of Pop III stellar evolutionary tracks \citep[][]{Yoon12,Costa25,Hassan25}, it should be noted that not all models predict this \citep[e.g.][]{Windhorst18}. 

\subsection{Dust obscuration in young star clusters}
The results presented in Sections~\ref{sec:normal star cluster} and \ref{sec:extreme IMF} suggest that mirror images with discrepant SED are expected to be detectable at high redshifts only in the case of very young star clusters ($\lesssim 5$ Myr). This overlaps with the age range over which young star clusters in the local Universe exhibit substantial dust obscuration, sometimes corresponding to several magnitudes of optical attenuation \citep[e.g.][]{Larsen10,Knutas25}. This obscuration would negatively impact our ability to detect lensed star clusters at sufficiently low total masses ($\approx 10^4$--$3\times 10^4\ M_\odot$ in the case of a standard IMF; Figure~\ref{fig:average coldiff}) to display the discrepant mirror-image SED phenomenon, but whether the obscuration is equally prominent at the low metallicites that may be relevant for the high-redshift Universe remains an open question.
 
\subsection{Nebular emission}
The results presented in this paper relate to the stellar component of lensed star cluster SEDs only. However, at the star cluster ages where we argue that microlensing-induced differences between mirror-image SEDs occur ($\lesssim 5$ Myr), a very significant part of the rest-frame ultraviolet and optical SED may come from nebular emission (nebular continuum and emission lines; e.g. \citealt{Zackrisson13}) due to ionizing radiation from hot stars being reprocessed in the interstellar medium. If this nebular light spatially blends with that of the stars, as is usually the case of unresolved, unlensed high-redshift galaxies, then the ability to detect microlensing-induced SED differences between star cluster images would be diminished. Since the nebular region of a young star cluster is several orders of magnitude larger than even the most extreme stars in the star cluster, we do not expect the nebular light to display temporal brightness fluctuations due to microlensing. Hence, nebular light dampens the variations in the overall star cluster SED by adding a non-varying flux component, provided that the stars and nebula spatially coincide in the observational data. Whether nebular and stellar light will blend completely at the magnifications we consider ($\mu\sim 100$) is non-trivial to predict, because of the large uncertainties in the detailed response of the surrounding gas from feedback from the stars as a function of time. However, a casual estimate (which neglects the mechanical feedback from stellar winds and supernovae, the latter of which may kick in at 2--3 Myr for the most massive stars and aid in blowing the gas out to large distances) suggests that a substantial fraction of the light from nebula may blend into the diffuse light of the arc rather than being completely mixed with that of the more point-like star cluster. The HII-region of newborn star clusters is expected to start off extremely dense and compact, but already after $\sim $0.1--1 Myr expand to scales of several parsec \citep[e.g.][]{Tremblin14,Sibony22}, i.e., larger than the effective radii of our fiducial star clusters. Young, low-metallicity $10^4$--$10^5\ M_\odot$ star clusters with standard IMFs can maintain a ionizing-photon production rate of $\sim 10^{51}$--$10^{52}$ s$^{-1}$ \citep[e.g.][]{Leitherer99}, which at a typical HII-region density of 100 cm$^{-2}$ would produce an HII region with a Str\"o{}mgren sphere radius of $\approx 15$--30 pc. If centered on a star cluster at 10 pc from the caustic, strong differential magnification would occur across this nebula, but for an assumed average tangential magnification of $\mu_\mathrm{t}\approx 100$, this light would be stretched across $\approx 0.5$--1 arcsec, i.e.  a factor of $\approx 4$--25 times the FWHM of the JWST/NIRcam point-spread function ($\approx 0.04$--0.14 arcsec in the $\approx 1$--4.4 $\mu$m range). Hence, complete confusion of stellar and nebular light is unlikely to occur. Moreover, if the nebular light appears sufficiently extended, it may be possible to subtract this component from that of the unresolved star cluster macroimage and thereby obtain an approximation of the purely stellar SED.

\subsection{Binary stars and correlated microlensing magnifications}
\label{subsec: binary stars}
Our current modelling is based on the assumption that the microlensing magnifications of the stars within a star cluster are uncorrelated, in the sense that the microlensing magnification of one star cluster member is expected to be independent from that of its closest neighbour. This assumption breaks down for small projected separations between stars, since the effects of microlensing are correlated on microarcsecond scales in the source plane (see Figures 1 and 4 in \citealt{Palencia24} for examples featuring both single microcaustics and dense networks of such caustics). Hence, two stars located at projected distances comparable to this scale (1 microarcsecond is $\approx 6\times 10^{-3}$ pc for $z_\mathrm{s}=6$) would experience correlated magnifications within the same mirror image. Under the assumption that all stars are single, the high-mass members of star clusters modelled in this paper will typically not attain such small separations  -- 1000 high-mass stars (a reasonable estimate for $M\gtrsim 10\ M_\odot$ stars in a $10^5\ M_\odot$ star cluster, i.e. the most massive star cluster we consider) randomly distributed within a 1 pc circle would get a median separation of $\approx 3\times 10^{-2}$ pc, i.e. $\approx 5$ microarcseconds. However, the binary fraction among the most massive stars in the local Universe is close to unity \citep[e.g.][]{Offner23}, with median separations $\lesssim 10^2$ AU, which would correspond to $\lesssim 0.1$ microarcseconds at $z_\mathrm{s}=6$. If high-mass stars in the early Universe exhibit binary properties similar to those observed the present-day Universe, correlated microlensing magnifications between the members of each binary pair are therefore expected.

What impact would this have on microlensing-induced differences between the SEDs star cluster macroimages? A detailed treatment is beyond the scope of this paper, but a general expectation is that the typical colour difference between mirror images ($|\Delta(m_{115}-m_{444})|$) of a star cluster with a standard IMF would increase if the brightest red stars typically form binary pairs with each other, but decrease if they typically team up with one of the brightest blue stars. To see why, consider a young star cluster in the state where a few of the most massive stars have evolved into $T_\mathrm{eff}<10000$ K giants (the state where mirror-image SED discrepancies are the most conspicuous, according to our models). At this point, the flux on the short-wavelength part ($m_{115}$) of our $z=6$ star cluster SEDs tends to be dominated by contributions from a large number of blue, main-sequence stars, whereas the long-wavelength side ($m_{444}$) is dominated by just a small number of red, evolved stars (Appendix~\ref{sec:appendix_light_contribution}). In the local Universe, most binaries involving bright, blue stars (spectral type O and B) as the higher-mass primary tend to display mass ratios (as quantified by $q=M_2/M_1$, where $M_1$ is the mass of the primary star) higher than expected from random pairing \citep{Offner23}\footnote{except at separation $>10^2$ AU (see Table 1 in \citealt{Offner23}), but this is a subdominant group of known OB-star binaries}. This implies that most of the blue, bright stars will team up with other blue stars of somewhat lower mass and luminosity. Since the microlensing magnification of these blue binaries are coupled, this effectively reduces the number of independently varying blue light sources within a star cluster image, and increases the overall microlensing variability in the short-wavelength part of the SED. In the limit of $q\approx1$, this boost in variability would be significant and could potentially serve as a probe of the binary population \citep{Dai21}. However, what dominates the variability in {\it colour} of a star cluster image is to what extent the red and blue sides of the SED are coupled. If the red stars form binaries with each other (red-red pairs), then the long-wavelength side of the star cluster SED should vary even more than the blue side (due to the smaller number of red stars overall). The red and blue sides of the SED would moreover vary almost independently of each other, thereby boosting both the colour variations within a star cluster image and the colour difference $|\Delta(m_{115}-m_{444})|$ between images. If, on the other hand, the brightest red stars will team up with some of the brightest blue stars, the opposite will happen. When the red part of the star cluster SED in one of the macroimages is brightening (or fading) due to a high (low) microlensing magnification, then the blue side will also tend to brighten (fade) due to similar microlensing boost of its blue companion, and colour differences between the macroimages will be dampened. In observational samples, low-metallicity red supergiants in binary systems do tend to team up with blue companions, although the binary fraction for red supergiants is much lower ($\approx 20\%$) than that of massive stars in general, and with a dearth of high$-q$ companions for the most massive red supergiants \citep{Patrick22}. Hence, detailed modelling of the binary-star population of star clusters would be required to explore how colour differences between mirror images of star clusters are affected by realistic binary systems.

\subsection{Mass segregation and very high surface mass densities in star clusters}
Simulations \citep[e.g.][]{Polak25} and observations \citep[e.g.][]{Khorrami21} of young, $M_\star \sim 10^3$--$10^5\ M_\odot$ star clusters support the notion that such objects can develop mass segregation, with an excess of high-mass stars in the cluster centre compared to the outskirts, on time scales $\leq 3$ Myr. A pile-up of high-mass stars with short projected separations would compound the effects of binary stars in violating the assumption of uncorrelated microlensing variations between stars in the same macroimage. However, in order for mass segregation to give rise to typical projected separations among $M>10\ M_\odot$ stars at the level where correlated microlensing magnifications are expected ($\lesssim 1$ microarcsecond; $\leq 6\times 10^{-3}$ pc at $z_\mathrm{s}=6$; see Section~\ref{subsec: binary stars}) would require such stars to be placed closer together than observed for most of the high-mass stars in the inner parsec of the young and mass-segregated star cluster R136 \citep{Khorrami21}.

Star clusters with extremely high surface mass densities could also result in stellar separations within the correlated-microlensing regime. For a stellar mass distribution that reflects the \citet{Kroupa01} IMF at all radii (i.e. is not heavily mass segregated), this happens at surface mass densities $\gtrsim 10^6\ M_\odot$ pc$^{-2}$. Such densities have indeed been claimed for the most massive ($\gtrsim 10^7\ M_\odot$) nuclear star clusters in  the low-redshift Universe \citep{Neumayer20}. The densest $z\gtrsim 6$ star clusters detected so far have average surface mass densities within the half-mass radius of $10^5$--$10^6\ M_\odot$ pc$^{-2}$ \citep[e.g.][]{Claeyssens26} -- with potentially even higher central surface densities -- but these are also more massive ($\sim 10^6\ M_\odot$) than the star clusters for which we argue that mirror images with discrepant SEDs would be detectable ($<10^5\ M_\odot$).

\subsection{Microlensing surface mass density}
Throughout this paper, we have assumed a surface mass density in microlenses of $\Sigma_\star=10\ M_\odot$ pc$^{-2}$ in the vicinity of the star cluster macroimages.  Both higher and lower values are in principle possible, depending on the position of the macroimages with respect to the distribution of intracluster light and galaxies within the galaxy cluster lens. What effect does the $\Sigma_\star$ parameter have on the discrepant-SED phenomenon? As shown in Figures 5 and 7 in \citet{Palencia24}, the shape of the probability density distributions for microlensing magnifications depend strongly on $\Sigma_\mathrm{eff}/\Sigma_\mathrm{c}$, i.e. the ratio of the effective microlensing surface mass density ($\Sigma_\mathrm{eff}=\mu_\mathrm{t}\Sigma_\star$) to the critical surface mass density for lensing. In their figures, the probability distributions undergo very complicated evolution as $\Sigma_\mathrm{eff}/\Sigma_\mathrm{c}$ is increased from 0.01 to 66.50, but the salient point relevant to our current study is that the magnification distributions of both the positive- and negative-parity image become flatter and less concentrated as $\Sigma_\mathrm{eff}/\Sigma_\mathrm{c}$ is increased up to $\Sigma_\mathrm{eff}/\Sigma_\mathrm{c}\approx 0.5$, and then start to become more concentrated again at higher values. Flatter magnification distributions act to increase the chance of a dominant star attaining disparate magnifications in the two macroimages, which boost the chance of having the overall star cluster SED displaying a different red/blue balance in the two images. Hence, one would expect the discrepant-SED effect to be maximized around $\Sigma_\mathrm{eff}/\Sigma_\mathrm{c}\approx 0.5$. For our fiducial case of $\Sigma_\star=10\ M_\odot$ pc$^{-2}$, $z_\mathrm{s}=6$ and $\mu_t\approx 120$, $\Sigma_\mathrm{eff}=\mu_\mathrm{t}\Sigma_\star\approx 0.7$, which is not far from this maximum. Based on \citet{Palencia24}, we do not expect substantial changes to our results in the case of factor of $\approx 2$ variations around our fiducial $\Sigma_\star = 10\ M_\odot$ pc$^{-2}$. However, factors of ten up or down (i.e. $\Sigma_\star \sim 1$ or $\sim 100\ M_\odot$ pc$^{-2}$) would both lead to more strongly peaked magnification distributions and hence lower median colour variations $|\Delta(m_{115}-m_{444})|$ between macroimages, which we have numerically verified using M$\_$SMiLe.
 
\subsection{Redshift effects on JWST/NIRCam observations}
A condition for seeing substantial SED differences between mirror-images of lensed star clusters observed using JWST NIRCam photometry is that the imprint of blue, main-sequence stars can be captured in the short-wavelength filters, while yellow/red supergiants remain detectable in at least one of the longest wavelength filters. This requirement is violated both at $z\lesssim 1$ and at $z\gtrsim 10$. At $z\lesssim 1$, NIRCam will be limited to probing the rest-frame optical (wavelengths  $\gtrsim 0.35\ \mu$m in the F070W filter), where high-mass main sequence star will appear very faint, and at $z\gtrsim 10$, the SED peak of yellow/red supergiants will have redshifted far outside the NIRCam wavelength range, making their contribution to even the longest-wavelength NIRCam filter (F480M) insignificant. JWST/MIRI could in principle be used to capture these low-$T_\mathrm{eff}$ stars at $z\gtrsim 10$, but the lower sensitivity of MIRI compared to NIRCam makes this prohibitive.

\subsection{IMF sampling}
Throughout this paper, we have assumed that star clusters are subject to random (or stochastic) IMF sampling, in contrast to the method known as optimal IMF sampling \citep[see][for a discussion on these different sampling assumptions]{Yan23}. In the random-sampling case, a $10^2\ M_\odot$ star cluster could in principle form just a single star of mass $10^2\ M_\odot$ (although with very low probability), whereas the optimal-sampling assumption would place an upper limit at $\sim 10\ M_\odot$ for the most-massive star within a $10^2\ M_\odot$ star cluster. The distinction is important for $\leq 10^4\ M_\odot$ star clusters, as it affects both the maximum and average mass of stars within such systems \citep{Kroupa13}, and hence their early SED evolution. Enforcing optimal IMF sampling on the least massive star clusters we consider ($3\times 10^3$--$1\times 10^4\ M_\odot$) would likely result in lower intrinsic brightness and lower probability for large mirror-image SED discrepancies, as fewer of the highest-mass stars would be able to form. Given the recent evidence from local star clusters in favour of optimal-sampling models \citep{Yan23,Chavez25}, it would in the future be interesting to consider such scenarios further.

\subsection{Impact of dark matter}
Throughout this paper, we have implicitly assumed that the microlensing that is affecting the mirror images of our modelled high-redshift star clusters is caused solely by stars in the intracluster medium of the foreground lensing cluster. However, as the critical curves for high-redshift sources are located at larger projected distances from the centre of the lensing cluster compared to the critical curves for low-redshift sources, one may expect that the relative contribution from any putative compact dark matter components to the microlensing effect may increase with increasing source redshift. Hence, effects of the type discussed here could serve as an indirect probe of compact dark matter objects like primordial black holes \citep[e.g.][]{Diego18}. Diffuse cold dark matter subhalos  can also affect where microlensing is expected to happen and how it will manifest itself \citep[e.g.][]{Diego24,Ji25}.

\subsection{Momentary colour differences versus temporal variability}
Discrepancies in the SEDs of star cluster mirror images serve as indications that some of the brightest stars within the cluster are strongly affected by gravitational microlensing. However, microlensing also cause the fluxes of these images to vary over time \citep{Dai21}. The two signatures are complementary, but require different data sets for detection. Colour differences between mirror images can be detected in single-epoch, multi-filter data, wheres temporal variations require multiple epochs in or (or more) filters. Most of the JWST observations of cluster-lens fields are currently of the multi-filter, single-epoch type (or with multiple epochs observed with different filters), since such data sets serve a broader range of science goals. However, it is possible that several epochs of same-filter photometric data could be accumulated for a significant fraction of these cluster-lens field throughout the lifetime of JWST. 

The variability signature does have some notable advantages over the momentary colour-discrepancy signature. If there are Pop III star clusters with very top-heavy IMFs and extremely massive stars, then the colour-discrepancy method can only catch this at favourable ages where very bright, red stars make a sizeable contribution to the long-wavelength part of the SED, and only for combinations of filters and redshifts where this contribution is detectable. Such star clusters would also exhibit detectable temporal variability across the whole SED. This variability would be observable over a broader range of redshifts and star cluster ages, and would not hinge on whether the highest-mass Pop III stars evolve to become red supergiants at the end of their lifetimes.

\section{Summary}
\label{sec:summary} 
We have made the case that the images of macrolensed, high-redshift star clusters seen across a critical curve in some situations may exhibit different SED shapes. This happens because bright stars within the star clusters can experience different amounts of microlensing in the two macroimages. In most cases, the colour differences between the macroimages will be too small to be detected in typical JWST observations, but very young ($\lesssim 5$ Myr) star clusters at masses $< 10^5\ M_\odot$ are predicted to exhibit conditions suitable for manifesting conspicuous colour differences. The conditions required for seeing the effect are that: 1) the star cluster attains an overall brightness sufficient to be detected at the typical magnifications ($\sim 100$) of lensed star cluster images; 2) some of the most massive stars have evolved into the $T_\mathrm{eff}<20000$ K regime; and 3) some part of the SED is dominated by just a handful of stars. At combinations of star cluster masses and ages where these conditions are met, the median colour difference between the macroimages becomes larger if the star cluster IMF is very top-heavy and extends to extremely high masses, as may be the case for Pop III star clusters. A search for mirror images with mismatched SEDs in single-epoch surveys of strongly lensed cluster lens fields may therefore be worthwhile to identify candidates for Pop III star clusters with top-heavy IMFs in the early Universe. 

\section*{Acknowledgements}
EZ acknowledges project grant 2022-03804 from the Swedish Research Council (Vetenskapsr\aa{}det). EZ and FG acknowledge financial support from the Carl Trygger Foundation for scientific research (grant CTS 24: 3297). GC acknowledges partial financial support from European Union—Next Generation EU, Mission 4, Component 2, CUP: C93C24004920006, project ‘FIRES'. The referee is thanked for very useful comments on the manuscript, which helped improve the quality of the paper.


\section*{Data Availability}
The models used in this paper are available from the lead author upon reasonable request.



\bibliographystyle{mnras}
\bibliography{references} 

\appendix
\section{Light contribution to different parts of the JWST/NIRCam SED}
\label{sec:appendix_light_contribution}
\begin{figure*}
    \includegraphics[width=\columnwidth]{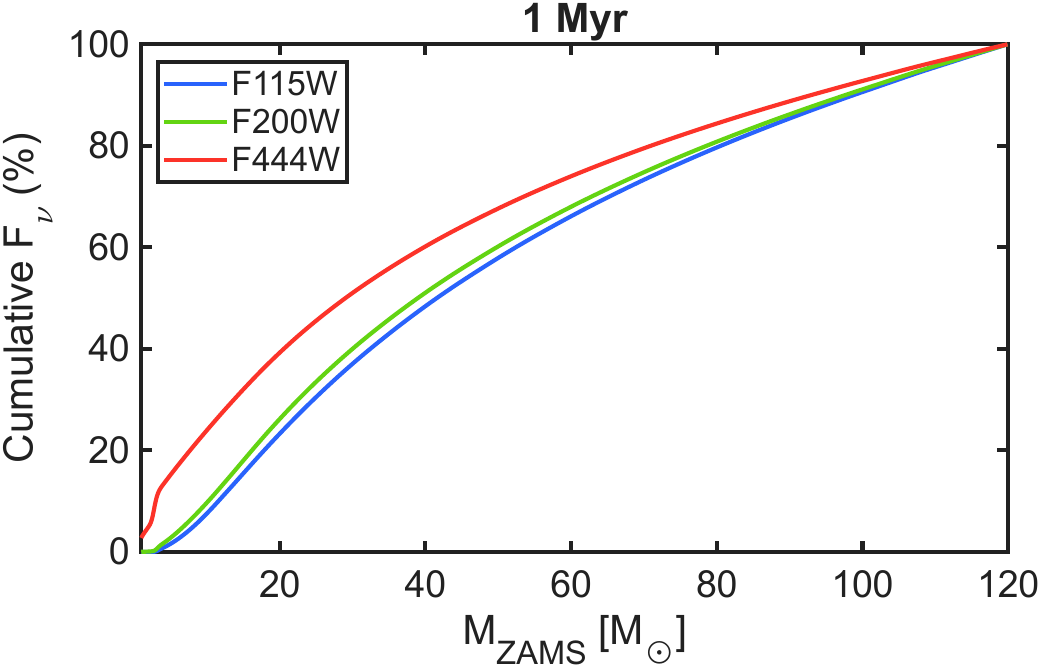}
    \includegraphics[width=\columnwidth]{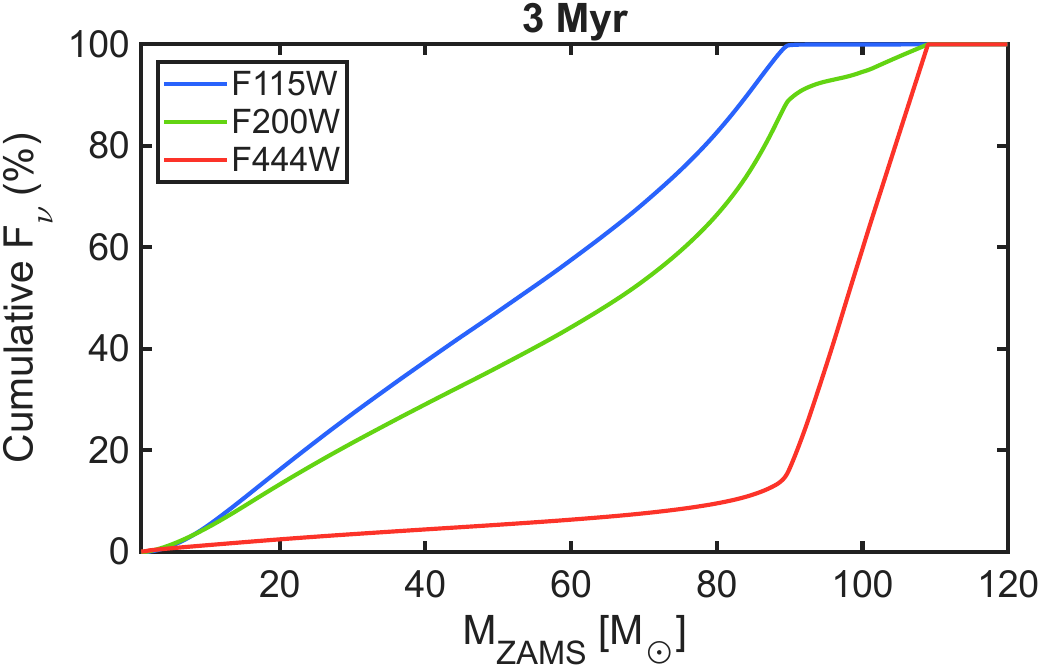}
    \caption{Relative contributions from stars of different masses to the integrated light in the JWST/NIRCam F115W (blue), F200W (green) and F444W (red) filters, in the case of a $z=6$, $Z=0.002$ star cluster with a standard, fully-sampled IMF, at ages 1 Myr (left) and 3 Myr (right)}.
\label{fig:A1}		
\end{figure*}

Here, we illustrate the relative contribution from stars of different masses to the integrated JWST/NIRCam SED in our modelled star clusters. Figure~\ref{fig:A1} shows the situation at ages 1 and 3 Myr in the case of our fiducial $z=6$ star clusters with standard, fully-sampled IMFs. These two ages exemplify the situations described in the example in Section~\ref{subsec:differential_magnification_example}, i.e. one case where all high-mass stars still have $T_\mathrm{eff}>30000$ K (1 Myr), and one where a few of the most massive ones have evolved to lower $T_\mathrm{eff}$ (3 Myr). At 1 Myr, stars over a wide range of masses contribute significantly to the integrated flux across the whole NIRCam SED (here illustrated by the F115W, F200W and F444W filters). At 3 Myr, on the other hand, most of the light ($\approx 90\%$) in the F444W filter comes from a small mass range ($\approx 90-110\ M_\odot$, with $M\gtrsim 110\ M_\odot$ stars having expired by this age).

\section{Pop III star clusters with a log-flat IMF and 2--300 $M_\odot$ stars}
\label{sec:appendix_PopIII_2_300_logflat}
\begin{figure*}
	\includegraphics[width=\columnwidth]{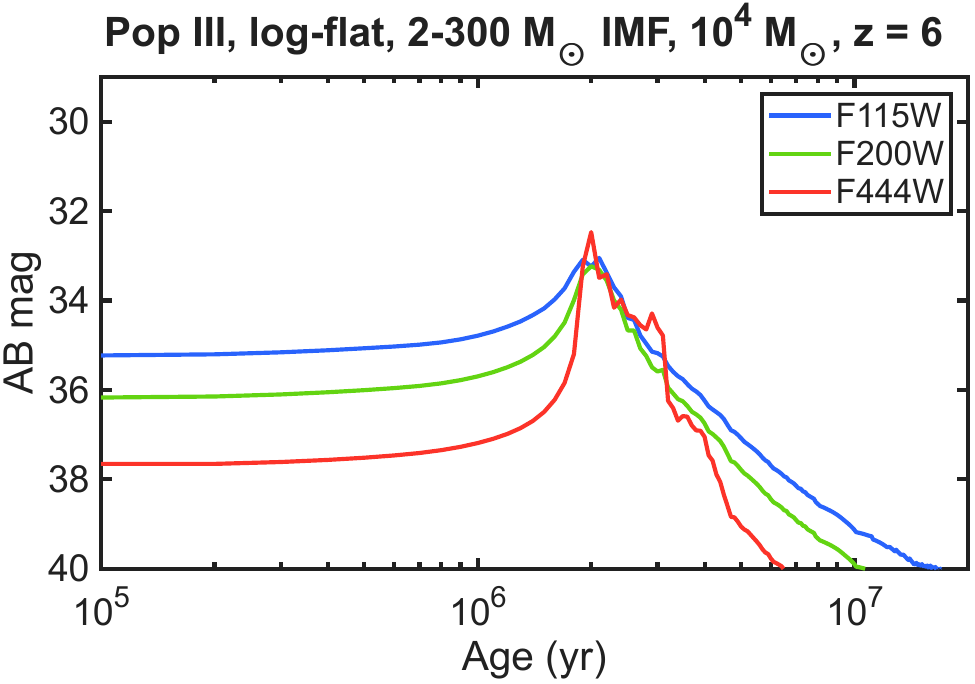}
    \includegraphics[width=\columnwidth]{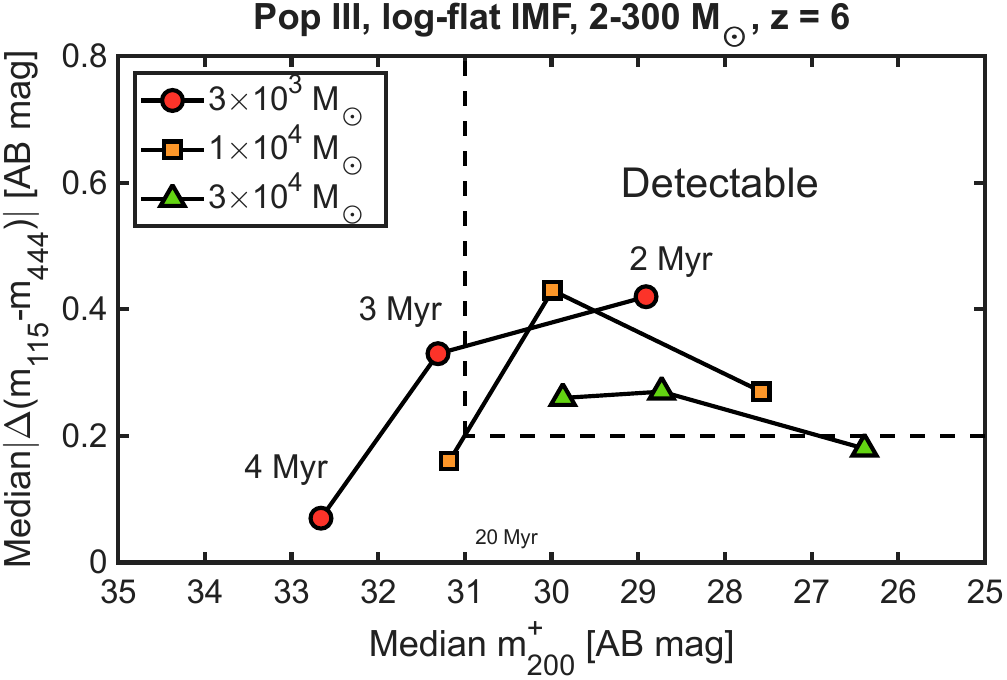}
    \caption{Pop III star clusters with a log-flat IMF slope and mass range 2--300 $M_\odot$. {\bf Left:} Same as Figure~\ref{fig:PopIII_1e4_colevol}, but with upper mass limit 300 $M_\odot$  {\bf Right:} Same as the right panel of Figure~\ref{fig:PopIII_example_and_average_coldiff}, but with upper mass limit 300 $M_\odot$ and ages of 2, 3 and 4 Myr shown.}
\label{fig:B1}		
\end{figure*}

Here, we explore the consequences of altering the upper stellar mass limit from $2000\ M_\odot$ (as in Section~\ref{sec:extreme IMF}) to 300 $M_\odot$ for our Pop III star cluster models, while retaining the same top-heavy (log-flat) IMF slope. In Figure~\ref{fig:B1},  we show figures corresponding to Figure~\ref{fig:PopIII_1e4_colevol} and the right panel of Figure~\ref{fig:PopIII_example_and_average_coldiff}, but with upper stellar mass 300 $M_\odot$.  Section~\ref{sec:extreme IMF}. In the 2-300 $M_\odot$ case, low-$T_\mathrm{eff}$ stars become prominent in the star cluster SED once the star cluster has reached an age of $\approx 2$ Myr, i.e. somewhat later than in the 2--2000 $M_\odot$ case ($\approx 1.5$ Myr), because the most massive stars with the shortest lifetimes are now excluded. At fixed star cluster mass, the 2--300 $M_\odot$ model is brighter in F200W at ages 2--3 Myr  than the 2--2000 $M_\odot$ model, since the 2--300 $M_\odot$ model has a larger mass fraction locked up in stars that remain alive at that age. For star clusters $1\times 10^4\ M_\odot$ and $3\times 10^4\ M_\odot$, detectable median $|\Delta(m_{115}-m_{444})|$ colour differences are produced during a period of just under 2 Myr, and at $3\times 10^3\ M_\odot$ during just under 1 Myr. The $|\Delta(m_{115}-m_{444})|$ produced at $m_{200}^{+}\leq 28$ AB mag are smaller than in the 2--2000 $M_\odot$ case, since a greater total number of stars contribute significantly to the light of the star cluster in the 2--300 $M_\odot$ case in this brightness range. These results are in accordance with the discussion on the role of the most massive stars presented in Section~\ref{sec:extreme IMF}.

\section{Effects of source size on the magnification probability distribution}
\label{sec:appendix_source_size}
\begin{figure*}
	\includegraphics[width=\columnwidth]{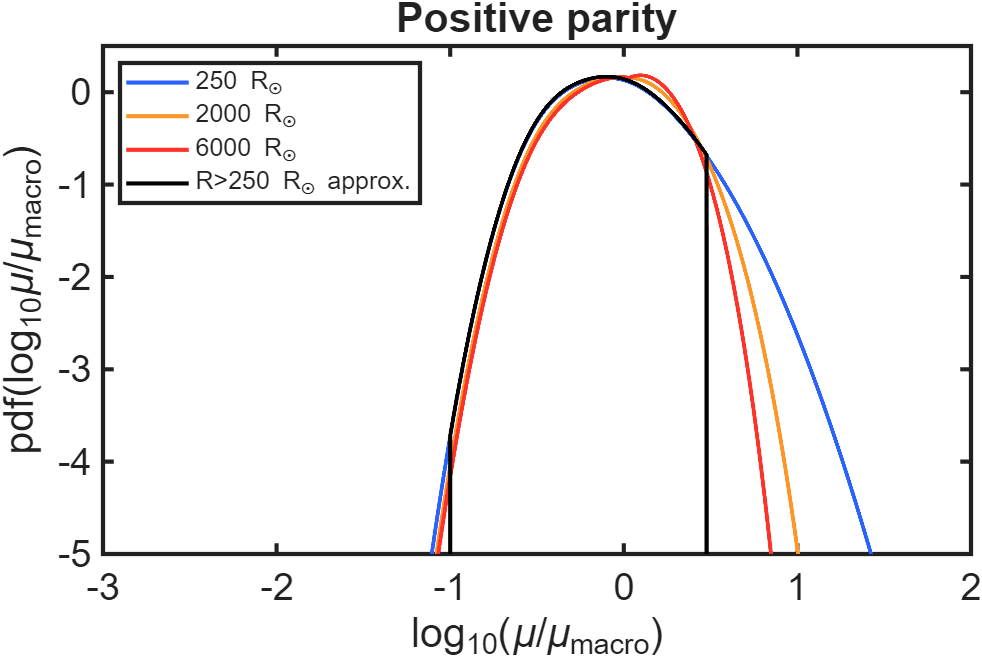}
    \includegraphics[width=\columnwidth]{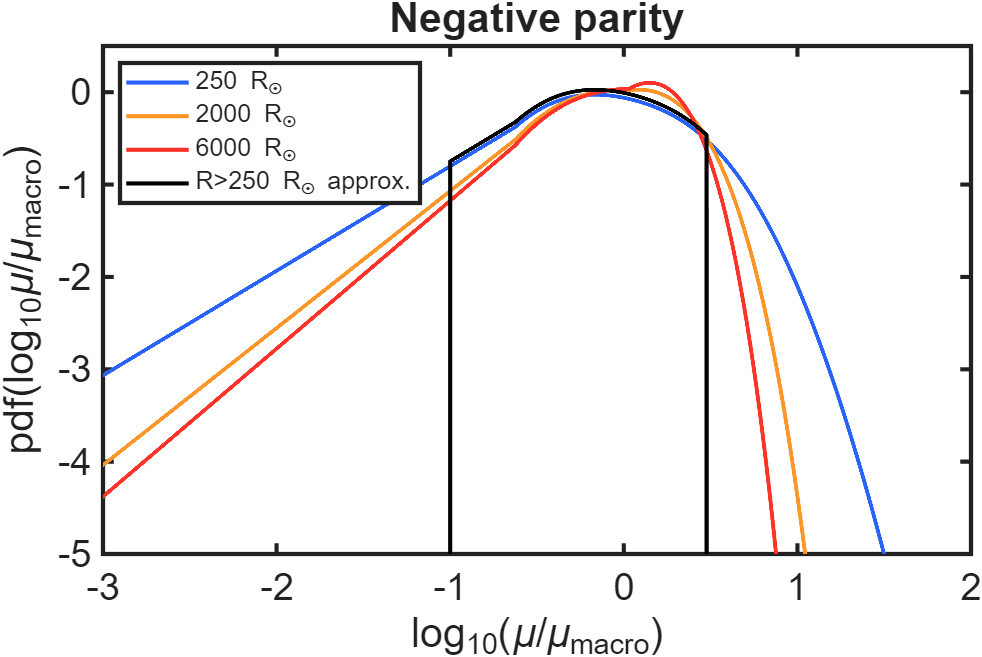}
    \caption{Probability of magnification (in $\log_{10} \mu/\mu_\mathrm{macro}$ bins) for positive-parity (left) and negative-parity (right) images of sources with different source radii, in the case of $\Sigma_\star=10$ $M_\odot$ pc$^{-2}$ and $z_\mathrm{s}=6$. The blue line (source radius 250 $R_\odot$) represents the standard M{\_}SMiLe output, whereas the orange ($\approx 2000\ R_\odot$) and red ($\approx6000\ R_\odot$) lines have been generated with a top-hat kernel that mimics larger source sizes. The black line represents the conservative $R>250\ R_\odot$ approximation used in this paper. Here, $\mu_\mathrm{macro}$ represents the macromagnification of a single macroimage ($\mu_\mathrm{macro}=\mu_\mathrm{t}\mu_\mathrm{r}$).}
\label{fig:C1}		
\end{figure*}

In Figure~\ref{fig:C1}, we illustrate the approximation used to limit the magnification range in the case of source sizes larger than the native M{\_}SMiLe resolution.

By convolving \citet{Palencia24} magnification maps with circular top-hat kernels to mimic different source sizes, we show examples of the predicted magnification probability distribution for sources with radius $\approx 2000\ R_\odot$ and $\approx 6000 R_\odot$, as compared to those generated with the default resolution ($250\ R_\odot$), in the case of $z_\mathrm{s}=6$ and $\Sigma_\star=10$ $M_\odot$ pc$^{-2}$. The larger source radii reduce the probability for very high magnifications (in both the positive- and negative-parity images) but also the probability for very small magnifications (primarily in negative-parity images). The sharp truncation beyond the 0.1--3$\mu_\mathrm{t}\mu_\mathrm{r}$ range used for $R>250\ R_\odot$ in this paper is conservative, in the sense that it systematically underpredicts the probability for large deviations from the macromagnification ($\mu_\mathrm{t}\mu_\mathrm{r}$) of an image. In the examples shown, the probability for magnifications outside the 0.1--3$\mu_\mathrm{t}\mu_\mathrm{r}$ range is only $\approx 0.1\%$ ($\approx 3\%$) for the positive (negative) parity images of a $R=6000\ R_\odot$ stars. For $R=2000\ R_\odot$ stars, the corresponding probabilities are $\approx 2\%$ ($\approx 6\%$). While Figure~\ref{fig:C1} reveals that there are other source-size effects that our sharp truncation fails to capture (gradual decline in probability in negative-parity images just above $0.1\mu_\mathrm{t}\mu_\mathrm{r}$ and a bump of enhanced probability just above $\mu_\mathrm{t}\mu_\mathrm{r}$), our overall expectation is that the sharp truncation should result in conservative estimates of the size of typical SED differences between the mirror images of star clusters, in the case where these differences are dominated by very large stars. Tests using the default M{\_}SMiLe magnification probability distributions even for large stars (without any truncation beyond 0.1--3$\mu_\mathrm{t}\mu_\mathrm{r}$ at $R>250\ R_\odot$) in the case of the 3 Myr old, $3\times 10^4\ M_\odot$, $Z=0.002$ star clusters in Figure~\ref{fig:average coldiff} support this notion, as this increases $|\Delta(m_{115}-m_{444})|$.

\bsp	
\label{lastpage}
\end{document}